\documentstyle[graphicx]{mn2e}
\newcommand{\omhh}{\Omega_{\rm m} h^2}
\newcommand{\obhh}{\Omega_{\rm b} h^2}
\newcommand{\om}{\Omega_{\rm m}}
\newcommand{\ob}{\Omega_{\rm b}}
\newcommand{\ok}{\Omega_{\rm k}}
\newcommand{\ol}{\Omega_\Lambda}
\begin{document}

\title[WiggleZ Survey: BAOs at $z=0.6$]{The WiggleZ Dark Energy
  Survey: testing the cosmological model with baryon acoustic
  oscillations at $z=0.6$}

\author[Blake et al.]{\parbox[t]{\textwidth}{Chris
    Blake$^1$\footnotemark, Tamara
    Davis$^{2,3}$, Gregory B.\ Poole$^1$, David Parkinson$^2$, \\
    Sarah Brough$^4$, Matthew Colless$^4$, Carlos Contreras$^1$,
    Warrick Couch$^1$, \\ Scott Croom$^5$, Michael J.\ Drinkwater$^2$,
    Karl Forster$^6$, David Gilbank$^7$, \\ Mike Gladders$^8$, Karl
    Glazebrook$^1$, Ben Jelliffe$^5$, Russell J.\ Jurek$^9$, I-hui
    Li$^1$, \\ Barry Madore$^{10}$, D.\ Christopher Martin$^6$, Kevin
    Pimbblet$^{11}$, Michael Pracy$^{1,12}$, \\ Rob Sharp$^{4,12}$,
    Emily Wisnioski$^1$, David Woods$^{13}$, Ted K.\ Wyder$^6$ and
    H.K.C. Yee$^{14}$} \\ \\ $^1$ Centre for Astrophysics \&
  Supercomputing, Swinburne University of Technology, P.O. Box 218,
  Hawthorn, VIC 3122, Australia \\ $^2$ School of Mathematics and
  Physics, University of Queensland, Brisbane, QLD 4072, Australia \\
  $^3$ Dark Cosmology Centre, Niels Bohr Institute, University of
  Copenhagen, Juliane Maries Vej 30, DK-2100 Copenhagen \O, Denmark \\
  $^4$ Australian Astronomical Observatory, P.O. Box 296, Epping, NSW
  1710, Australia \\ $^5$ Sydney Institute for Astronomy, School of
  Physics, University of Sydney, NSW 2006, Australia \\ $^6$
  California Institute of Technology, MC 278-17, 1200 East California
  Boulevard, Pasadena, CA 91125, United States \\ $^7$ Astrophysics
  and Gravitation Group, Department of Physics and
  Astronomy, University of Waterloo, Waterloo, ON N2L 3G1, Canada \\
  $^8$ Department of Astronomy and Astrophysics, University of
  Chicago, 5640 South Ellis Avenue, Chicago, IL 60637, United States
  \\ $^9$ Australia Telescope National Facility, CSIRO, Epping, NSW
  1710, Australia \\ $^{10}$ Observatories of the Carnegie Institute
  of Washington, 813 Santa Barbara St., Pasadena, CA 91101, United
  States \\ $^{11}$ School of Physics, Monash University, Clayton, VIC
  3800, Australia \\ $^{12}$ Research School of Astronomy \&
  Astrophysics, Australian National University, Weston Creek, ACT
  2600, Australia \\ $^{13}$ Department of Physics \& Astronomy,
  University of British Columbia, 6224 Agricultural Road, Vancouver,
  BC V6T 1Z1, Canada \\ $^{14}$ Department of Astronomy and
  Astrophysics, University of Toronto, 50 St.\ George Street, Toronto,
  ON M5S 3H4, Canada}

\maketitle

\begin{abstract}
  We measure the imprint of baryon acoustic oscillations (BAOs) in the
  galaxy clustering pattern at the highest redshift achieved to date,
  $z=0.6$, using the distribution of $N=132{,}509$ emission-line
  galaxies in the WiggleZ Dark Energy Survey.  We quantify BAOs using
  three statistics: the galaxy correlation function, power spectrum
  and the band-filtered estimator introduced by Xu et al.\ (2010).
  The results are mutually consistent, corresponding to a $4.0\%$
  measurement of the cosmic distance-redshift relation at $z=0.6$ (in
  terms of the acoustic parameter ``$A(z)$'' introduced by Eisenstein
  et al.\ (2005) we find $A(z=0.6) = 0.452 \pm 0.018$).  Both BAOs and
  power spectrum shape information contribute toward these
  constraints.  The statistical significance of the detection of the
  acoustic peak in the correlation function, relative to a wiggle-free
  model, is $3.2$-$\sigma$.  The ratios of our distance measurements
  to those obtained using BAOs in the distribution of Luminous Red
  Galaxies at redshifts $z=0.2$ and $z=0.35$ are consistent with a
  flat $\Lambda$ Cold Dark Matter model that also provides a good fit
  to the pattern of observed fluctuations in the Cosmic Microwave
  Background (CMB) radiation.  The addition of the current WiggleZ
  data results in a $\approx 30\%$ improvement in the measurement
  accuracy of a constant equation-of-state, $w$, using BAO data alone.
  Based solely on geometric BAO distance ratios, accelerating
  expansion ($w < -1/3$) is required with a probability of $99.8\%$,
  providing a consistency check of conclusions based on supernovae
  observations.  Further improvements in cosmological constraints will
  result when the WiggleZ Survey dataset is complete.
\end{abstract}

\section{Introduction}
\renewcommand{\thefootnote}{\fnsymbol{footnote}}
\setcounter{footnote}{1}
\footnotetext{E-mail: cblake@astro.swin.edu.au}

The measurement of baryon acoustic oscillations (BAOs) in the
large-scale clustering pattern of galaxies has rapidly become one of
the most important observational pillars of the cosmological model.
BAOs correspond to a preferred length scale imprinted in the
distribution of photons and baryons by the propagation of sound waves
in the relativistic plasma of the early Universe (Peebles \& Yu 1970,
Sunyaev \& Zeldovitch 1970, Bond \& Efstathiou 1984, Holtzman 1989,
Hu \& Sugiyama 1996, Eisenstein \& Hu 1998).  A full account of the
early-universe physics is provided by Bashinsky \& Bertschinger (2001,
2002).  In a simple intuitive description of the effect we can imagine
an overdensity in the primordial dark matter distribution creating an
overpressure in the tightly-coupled photon-baryon fluid and launching
a spherical compression wave.  At redshift $z \approx 1000$ there is a
precipitous decrease in sound speed due to recombination to a neutral
gas and de-coupling of the photon-baryon fluid.  The photons stream
away and can be mapped as the Cosmic Microwave Background (CMB)
radiation; the spherical shell of compressed baryonic matter is frozen
in place.  The overdense shell, together with the initial central
perturbation, seeds the later formation of galaxies and imprints a
preferred scale into the galaxy distribution equal to the sound
horizon at the baryon drag epoch.  Given that baryonic matter is
secondary to cold dark matter in the clustering pattern, the amplitude
of the effect is much smaller than the acoustic peak structure in the
CMB.

The measurement of BAOs in the pattern of late-time galaxy clustering
provides a compelling validation of the standard picture that
large-scale structure in today's Universe arises through the
gravitational amplification of perturbations seeded at early times.
The small amplitude of the imprint of BAOs in the galaxy distribution
is a demonstration that the bulk of matter consists of non-baryonic
dark matter that does not couple to the relativistic plasma before
recombination.  Furthermore, the preferred length scale -- the sound
horizon at the baryon drag epoch -- may be predicted very accurately
by measurements of the CMB which yield the physical matter and baryon
densities that control the sound speed, expansion rate and
recombination time: the latest determination is $153.3 \pm 2.0$ Mpc
(Komatsu et al.\ 2009).  Therefore the imprint of BAOs provide a
standard cosmological ruler that can map out the cosmic expansion
history and provide precise and robust constraints on the nature of
the ``dark energy'' that is apparently dominating the current cosmic
dynamics (Blake \& Glazebrook 2003; Hu \& Haiman 2003; Seo \&
Eisenstein 2003).  In principle the standard ruler may be applied in
both the tangential and radial directions of a galaxy survey, yielding
measures of the angular diameter distance and Hubble parameter as a
function of redshift.

The large scale and small amplitude of the BAOs imprinted in the
galaxy distribution implies that galaxy redshift surveys mapping
cosmic volumes of order 1 Gpc$^3$ with of order $10^5$ galaxies are
required to ensure a robust detection (Tegmark 1997, Blake \&
Glazebrook 2003, Blake et al.\ 2006).  Gathering such a sample
represents a formidable observational challenge typically
necessitating hundreds of nights of telescope time over several years.
The leading such spectroscopic dataset in existence is the Sloan
Digital Sky Survey (SDSS), which covers $8000$ deg$^2$ of sky
containing a ``main'' $r$-band selected sample of $10^6$ galaxies with
median redshift $z \approx 0.1$, and a Luminous Red Galaxy (LRG)
extension consisting of $10^5$ galaxies but covering a
significantly-greater cosmic volume with median redshift $z \approx
0.35$.  Eisenstein et al.\ (2005) reported a convincing BAO detection
in the 2-point correlation function of the SDSS Third Data Release
(DR3) LRG sample at $z=0.35$, demonstrating that this standard-ruler
measurement was self-consistent with the cosmological model
established from CMB observations and yielding new, tighter
constraints on cosmological parameters such as the spatial curvature.
Percival et al.\ (2010) undertook a power-spectrum analysis of the
SDSS DR7 dataset, considering both the main and LRG samples, and
constrained the distance-redshift relation at both $z=0.2$ and
$z=0.35$ with $\sim 3\%$ accuracy in units of the standard ruler
scale.  Other studies of the SDSS LRG sample, producing broadly
similar conclusions, have been performed by Huetsi (2006), Percival et
al. (2007), Sanchez et al.\ (2009) and Kazin et al.\ (2010a).  Some
analyses have attempted to separate the tangential and radial BAO
signatures in the LRG dataset, albeit with lower statistical
significance (Gaztanaga et al.\ 2009, Kazin et al.\ 2010b).  These
studies built on earlier hints of BAOs reported by the 2-degree Field
Galaxy Redshift Survey (Cole et al.\ 2005) and combinations of smaller
datasets (Miller et al.\ 2001).

This ambitious observational program to map out the cosmic expansion
history with BAOs has prompted serious theoretical scrutiny of the
accuracy with which we can model the BAO signature and the likely
amplitude of systematic errors in the measurement.  The pattern of
clustering laid down in the high-redshift Universe is potentially
subject to modulation by the non-linear scale-dependent growth of
structure, by the distortions apparent when the signal is observed in
redshift-space, and by the bias with which galaxies trace the
underlying network of matter fluctuations.  In this context the fact
that the BAOs are imprinted on large, linear and quasi-linear scales
of the clustering pattern implies that non-linear BAO distortions are
relatively accessible to modelling via perturbation theory or
numerical N-body simulations (Eisenstein, Seo \& White 2007, Crocce \&
Scoccimarro 2008, Matsubara 2008).  The leading-order effect is a
``damping'' of the sharpness of the acoustic feature due to the
differential motion of pairs of tracers separated by 150 Mpc driven by
bulk flows of matter.  Effects due to galaxy formation and bias are
confined to significantly smaller scales and are not expected to cause
significant acoustic peak shifts.  Although the non-linear damping of
BAOs reduces to some extent the accuracy with which the standard ruler
can be applied, the overall picture remains that BAOs provide a robust
probe of the cosmological model free of serious systematic error.  The
principle challenge lies in executing the formidable galaxy redshift
surveys needed to exploit the technique.

In particular, the present ambition is to extend the relatively
low-redshift BAO measurements provided by the SDSS dataset to the
intermediate- and high-redshift Universe.  Higher-redshift
observations serve to further test the cosmological model over the
full range of epochs for which dark energy apparently dominates the
cosmic dynamics, can probe greater cosmic volumes and therefore yield
more accurate BAO measurements, and are less susceptible to the
non-linear effects which damp the sharpness of the acoustic signature
at low redshift and may induce low-amplitude systematic errors.
Currently, intermediate redshifts have only been probed by
photometric-redshift surveys which have limited statistical precision
(Blake et al.\ 2007, Padmanabhan et al.\ 2007).

The WiggleZ Dark Energy Survey at the Australian Astronomical
Observatory (Drinkwater et al.\ 2010) was designed to provide the
next-generation spectroscopic BAO dataset following the SDSS,
extending the distance-scale measurements across the
intermediate-redshift range up to $z = 0.9$ with a precision of
mapping the acoustic scale comparable to the SDSS LRG sample.  The
survey, which began in August 2006, completed observations in January
2011 and has obtained of order $200{,}000$ redshifts for UV-bright
emission-line galaxies covering of order 1000 deg$^2$ of equatorial
sky.  Analysis of the full dataset is ongoing.  In this paper we
report intermediate results for a subset of the WiggleZ sample with
effective redshift $z = 0.6$.

BAOs are a signature present in the 2-point clustering of galaxies.
In this paper we analyze this signature using a variety of techniques:
the 2-point correlation function, the power spectrum, and the
band-filtered estimator recently proposed by Xu et al.\ (2010) which
amounts to a band-filtered correlation function.  Quantifying the BAO
measurement using this range of techniques increases the robustness of
our results and gives us a sense of the amplitude of systematic errors
induced by our current methodologies.  Using each of these techniques
we measure the angle-averaged clustering statistic, making no attempt
to separate the tangential and radial components of the signal.
Therefore we measure the ``dilation scale'' distance $D_V(z)$
introduced by Eisenstein et al.\ (2005) which consists of two parts
physical angular-diameter distance, $D_A(z)$, and one part radial
proper-distance, $cz/H(z)$:
\begin{equation}
D_V(z) = \left[ (1+z)^2 D_A(z)^2 \frac{cz}{H(z)} \right]^{1/3} .
\label{eqdv}
\end{equation}
This distance measure reflects the relative importance of the
tangential and radial modes in the angle-averaged BAO measurement
(Padmanabhan \& White 2008), and reduces to proper distance in the
low-redshift limit.  Given that a measurement of $D_V(z)$ is
correlated with the physical matter density $\omhh$ which controls the
standard ruler scale, we extract other distilled parameters which are
far less significantly correlated with $\omhh$, namely: the acoustic
parameter $A(z)$ as introduced by Eisenstein et al.\ (2005); the ratio
$d_z = r_s(z_d)/D_V(z)$, which quantifies the distance scale in units
of the sound horizon at the baryon drag epoch, $r_s(z_d)$; and $1/R_z$
which is the ratio between $D_V(z)$ and the distance to the CMB
last-scattering surface.

The structure of this paper is as follows.  The WiggleZ data sample is
introduced in Section \ref{secdata}, and we then present our
measurements of the galaxy correlation function, power spectrum and
band-filtered correlation function in Sections \ref{secxi},
\ref{secpk} and \ref{secxu}, respectively.  The results of these
different methodologies are compared in Section \ref{seccomp}.  In
Section \ref{secdist} we state our measurements of the BAO distance
scale at $z=0.6$ using various distilled parameters, and combine our
result with other cosmological datasets in Section \ref{seccosmofit}.
Throughout this paper we assume a fiducial cosmological model which is
a flat $\Lambda$CDM Universe with matter density parameter $\om =
0.27$, baryon fraction $\ob/\om = 0.166$, Hubble parameter $h = 0.71$,
primordial index of scalar perturbations $n_{\rm s} = 0.96$ and
redshift-zero normalization $\sigma_8 = 0.8$.  This fiducial model is
used for some of the intermediate steps in our analysis but our final
cosmological constraints are, to first-order at least, independent of
the choice of fiducial model.

\section{Data}
\label{secdata}

\begin{figure}
\begin{center}
\resizebox{\columnwidth}{!}{\rotatebox{270}{\includegraphics{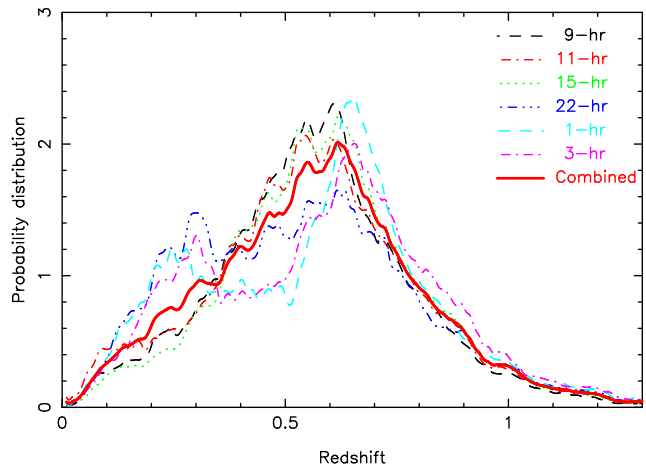}}}
\end{center}
\caption{The probability distribution of galaxy redshifts in each of
  the WiggleZ regions used in our clustering analysis, together with
  the combined distribution.  Differences between individual regions
  result from variations in the galaxy colour selection criteria
  depending on the available optical imaging (Drinkwater et
  al.\ 2010).}
\label{fignz}
\end{figure}

The WiggleZ Dark Energy Survey at the Anglo Australian Telescope
(Drinkwater et al.\ 2010) is a large-scale galaxy redshift survey of
bright emission-line galaxies mapping a cosmic volume of order 1
Gpc$^3$ over the redshift interval $z < 1$.  The survey has obtained
of order $200{,}000$ redshifts for UV-selected galaxies covering of
order 1000 deg$^2$ of equatorial sky.  In this paper we analyze the
subset of the WiggleZ sample assembled up to the end of the 10A
semester (May 2010).  We include data from six survey regions in the
redshift range $0.3 < z < 0.9$ -- the 9-hr, 11-hr, 15-hr, 22-hr, 1-hr
and 3-hr regions -- which together constitute a total sample of $N =
132{,}509$ galaxies.  The redshift probability distributions of the
galaxies in each region are shown in Figure \ref{fignz}.

The selection function for each survey region was determined using the
methods described by Blake et al.\ (2010) which model effects due to
the survey boundaries, incompleteness in the parent UV and optical
catalogues, incompleteness in the spectroscopic follow-up, systematic
variations in the spectroscopic redshift completeness across the
AAOmega spectrograph, and variations of the galaxy redshift
distribution with angular position.  The modelling process produces a
series of Monte Carlo random realizations of the angle/redshift
catalogue in each region, which are used in the correlation function
estimation.  By stacking together a very large number of these random
realizations we deduced the 3D window function grid used for power
spectrum estimation.

\section{Correlation function}
\label{secxi}

\subsection{Measurements}

The 2-point correlation function is a common method for quantifying
the clustering of a population of galaxies, in which the distribution
of pair separations in the dataset is compared to that within random,
unclustered catalogues possessing the same selection function (Peebles
1980).  In the context of measuring baryon acoustic oscillations, the
correlation function has the advantage that the expected signal of a
preferred clustering scale is confined to a single, narrow range of
separations around $105 \, h^{-1}$ Mpc.  Furthermore, small-scale
non-linear effects, such as the distribution of galaxies within dark
matter haloes, do not influence the correlation function on these
large scales.  One disadvantage of this statistic is that measurements
of the large-scale correlation function are prone to systematic error
because they are very sensitive to the unknown mean density of the
galaxy population.  However, such ``integral constraint'' effects
result in a roughly constant offset in the large-scale correlation
function, which does not introduce a preferred scale that could mimic
the BAO signature.

In order to estimate the correlation function of each WiggleZ survey
region we first placed the angle/redshift catalogues for the data and
random sets on a grid of co-moving co-ordinates, assuming a flat
$\Lambda$CDM model with matter density $\om = 0.27$.  We then measured
the redshift-space 2-point correlation function $\xi(s)$ for each
region using the Landy-Szalay (1993) estimator:
\begin{equation}
\xi(s) = \frac{DD(s) - DR(s) + RR(s)}{RR(s)} ,
\label{eqxiest}
\end{equation}
where $DD(s)$, $DR(s)$ and $RR(s)$ are the data-data, data-random and
random-random weighted pair counts in separation bin $s$, each random
catalogue containing the same number of galaxies as the real dataset.
In the construction of the pair counts each data or random galaxy $i$
is assigned a weight $w_i = 1/(1 + n_i P_0)$, where $n_i$ is the
survey number density [in $h^3$ Mpc$^{-3}$] at the location of the
$i$th galaxy, and $P_0 = 5000 \, h^{-3}$ Mpc$^3$ is a characteristic
power spectrum amplitude at the scales of interest.  The survey number
density distribution is established by averaging over a large ensemble
of random catalogues.  The $DR$ and $RR$ pair counts are determined by
averaging over 10 random catalogues.  We measured the correlation
function in 20 separation bins of width $10 \, h^{-1}$ Mpc between
$10$ and $180 \, h^{-1}$ Mpc, and determined the covariance matrix of
this measurement using lognormal survey realizations as described
below.  We combined the correlation function measurements in each bin
for the different survey regions using inverse-variance weighting of
each measurement (we note that this procedure produces an almost
identical result to combining the individual pair counts).

\begin{figure}
\begin{center}
\resizebox{\columnwidth}{!}{\rotatebox{270}{\includegraphics{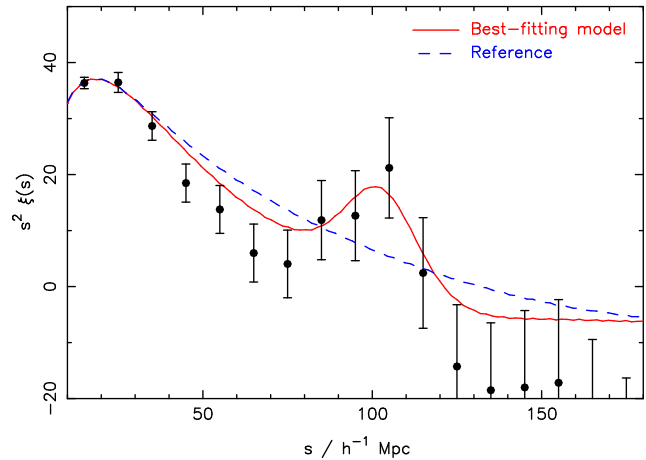}}}
\end{center}
\caption{The combined redshift-space correlation function $\xi(s)$ for
  WiggleZ survey regions, plotted in the combination $s^2 \, \xi(s)$
  where $s$ is the co-moving redshift-space separation.  The
  best-fitting clustering model (varying $\omhh$, $\alpha$ and $b^2$)
  is overplotted as the solid line.  We also show as the dashed line
  the corresponding ``no-wiggles'' reference model, constructed from a
  power spectrum with the same clustering amplitude but lacking baryon
  acoustic oscillations.}
\label{figxifit}
\end{figure}

The combined correlation function is plotted in Figure \ref{figxifit}
and shows clear evidence for the baryon acoustic peak at separation
$\sim 105 \, h^{-1}$ Mpc.  The effective redshift $z_{\rm eff}$ of the
correlation function measurement is the weighted mean redshift of the
galaxy pairs entering the calculation, where the redshift of a pair is
simply the average $(z_1+z_2)/2$, and the weighting is $w_1 w_2$ where
$w_i$ is defined above.  We determined $z_{\rm eff}$ for the bin $100
< s < 110 \, h^{-1}$ Mpc, although it does not vary significantly with
separation.  For the combined WiggleZ survey measurement, we found
$z_{\rm eff} = 0.60$.

We note that the correlation function measurements are corrected for
the effect of redshift blunders in the WiggleZ data catalogue.  These
are fully quantified in Section 3.2 of Blake et al.\ (2010), and can
be well-approximated by a scale-independent boost to the correlation
function amplitude of $(1-f_b)^{-2}$, where $f_b \sim 0.05$ is the
redshift blunder fraction (which is separately measured for each
WiggleZ region).

\subsection{Uncertainties : lognormal realizations and covariance
  matrix}

We determined the covariance matrix of the correlation function
measurement in each survey region using a large set of lognormal
realizations.  Jack-knife errors, implemented by dividing the survey
volume into many sub-regions, are a poor approximation for the error
in the large-scale correlation function because the pair separations
of interest are usually comparable to the size of the sub-regions,
which are then not strictly independent.  Furthermore, because the
WiggleZ dataset is not volume-limited and the galaxy number density
varies with position, it is impossible to define a set of sub-regions
which are strictly equivalent.

Lognormal realizations are relatively cheap to generate and provide a
reasonably accurate galaxy clustering model for the linear and
quasi-linear scales which are important for the modelling of baryon
oscillations (Coles \& Jones 1991).  We generated a set of
realizations for each survey region using the method described in
Blake \& Glazebrook (2003) and Glazebrook \& Blake (2005).  In brief,
we started with a model galaxy power spectrum $P_{\rm mod}(\vec{k})$
consistent with the survey measurement.  We then constructed Gaussian
realizations of overdensity fields $\delta_{\rm G}(\vec{r})$ sampled
from a second power spectrum $P_G(\vec{k}) \approx P_{\rm
  mod}(\vec{k})$ (defined below), in which real and imaginary Fourier
amplitudes are drawn from a Gaussian distribution with zero mean and
standard deviation $\sqrt{P_G(\vec{k})/2}$.  A lognormal overdensity
field $\delta_{\rm LN}(\vec{r}) = \exp{(\delta_{\rm G})} - 1$ is then
created, and is used to produce a galaxy density field $\rho_{\rm
  g}(\vec{r})$ consistent with the survey window function
$W(\vec{r})$:
\begin{equation}
\rho_{\rm g}(\vec{r}) \propto W(\vec{r}) \, [1 + \delta_{\rm
    LN}(\vec{r})] ,
\end{equation}
where the constant of proportionality is fixed by the size of the
final dataset.  The galaxy catalogue is then Poisson-sampled in cells
from the density field $\rho_{\rm g}(\vec{r})$.  We note that the
input power spectrum for the Gaussian overdensity field,
$P_G(\vec{k})$, is constructed to ensure that the final power spectrum
of the lognormal overdensity field is consistent with $P_{\rm
  mod}(\vec{k})$.  This is achieved using the relation between the
correlation functions of Gaussian and lognormal fields, $\xi_{\rm
  G}(\vec{r}) = \ln{[1 + \xi_{\rm mod}(\vec{r})]}$.

\begin{figure}
\begin{center}
\resizebox{\columnwidth}{!}{\includegraphics{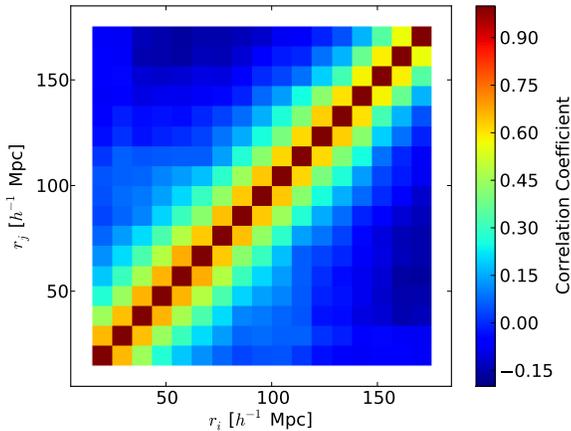}}
\end{center}
\caption{The amplitude of the cross-correlation $C_{ij}/\sqrt{C_{ii}
    C_{jj}}$ of the covariance matrix $C_{ij}$ for the correlation
  function measurement plotted in Figure \ref{figxifit}, determined
  using lognormal realizations.}
\label{figxicov}
\end{figure}

We determined the covariance matrix between bins $i$ and $j$ using the
correlation function measurements from a large ensemble of lognormal
realizations:
\begin{equation}
C_{ij} = \langle \xi_i \, \xi_j \rangle - \langle \xi_i \rangle
\langle \xi_j \rangle ,
\end{equation}
where the angled brackets indicate an average over the realizations.
Figure \ref{figxicov} displays the final covariance matrix resulting
from combining the different WiggleZ survey regions in the form of a
correlation matrix $C_{ij}/\sqrt{C_{ii} C_{jj}}$.  The magnitude of
the first and second off-diagonal elements of the correlation matrix
is typically $0.6$ and $0.4$, respectively.  We find that the
jack-knife errors on scales of $100 \, h^{-1}$ Mpc typically exceed
the lognormal errors by a factor of $\approx 50\%$, which we can
attribute to an over-estimation of the number of independent
jack-knife regions.

\subsection{Fitting the correlation function : template model and
  simulations}
\label{secxitemp}

In this Section we discuss the construction of the template fiducial
correlation function model $\xi_{\rm fid,galaxy}(s)$ which we fitted
to the WiggleZ measurement.  When fitting the model we vary a scale
distortion parameter $\alpha$, a linear normalization factor $b^2$ and
the matter density $\omhh$ which controls both the overall shape of
the correlation function and the standard ruler sound horizon scale.
Hence we fitted the model
\begin{equation}
\xi_{\rm mod}(s) = b^2 \, \xi_{\rm fid,galaxy}(\alpha \, s) .
\label{eqximod}
\end{equation}
The probability distribution of the scale distortion parameter
$\alpha$, after marginalizing over $\omhh$ and $b^2$, gives the
probability distribution of the distance variable $D_V(z_{\rm eff}) =
\alpha \, D_{V,{\rm fid}}(z_{\rm eff})$ where $z_{\rm eff} = 0.6$ for
our sample (Eisenstein et al.\ 2005, Padmanabhan \& White 2008).
$D_V$, defined by Equation \ref{eqdv}, is a composite of the physical
angular-diameter distance $D_A(z)$ and Hubble parameter $H(z)$ which
govern tangential and radial galaxy separations, respectively, where
$D_{V,{\rm fid}}(z_{\rm eff}) = 2085.4$ Mpc.

We note that the measured value of $D_V$ resulting from this fitting
process will be independent (to first order) of the fiducial
cosmological model adopted for the conversion of galaxy redshifts and
angular positions to co-moving co-ordinates.  A change in $D_{V,{\rm
    fid}}$ would result in a shift in the measured position of the
acoustic peak.  This shift would be compensated for by a corresponding
offset in the best-fitting value of $\alpha$, leaving the measurement
of $D_V = \alpha \, D_{V,{\rm fid}}$ unchanged (to first order).

An angle-averaged power spectrum $P(k)$ may be converted into an
angle-averaged correlation function $\xi(s)$ using the spherical
Hankel transform
\begin{equation}
\xi(s) = \frac{1}{2\pi^2} \int dk \, k^2 \, P(k) \left[
  \frac{\sin{(ks)}}{ks} \right] .
\label{eqpktoxi}
\end{equation}
In order to determine the shape of the model power spectrum for a
given $\omhh$, we first generated a linear power spectrum $P_L(k)$
using the fitting formula of Eisenstein \& Hu (1998).  This yields a
result in good agreement with a CAMB linear power spectrum (Lewis,
Challinor \& Lasenby 2000), and also produces a wiggle-free reference
spectrum $P_{\rm ref}(k)$ which possesses the same shape as $P_L(k)$
but with the baryon oscillation component deleted.  This reference
spectrum is useful for assessing the statistical significance with
which we have detected the acoustic peak.  We fixed the values of the
other cosmological parameters using our fiducial model $h = 0.71$,
$\obhh = 0.0226$, $n_{\rm s} = 0.96$ and $\sigma_8 = 0.8$.  Our
choices for these parameters are consistent with the latest fits to
the Cosmic Microwave Background radiation (Komatsu et al.\ 2009).

We then corrected the power spectrum for quasi-linear effects.  There
are two main aspects to the model: a damping of the acoustic peak
caused by the displacement of matter due to bulk flows, and a
distortion in the overall shape of the clustering pattern due to the
scale-dependent growth of structure (Eisenstein, Seo \& White 2007,
Crocce \& Scoccimarro 2008, Matsubara 2008).  We constructed our model
in a similar manner to Eisenstein et al.\ (2005).  We first
incorporated the acoustic peak smoothing by multiplying the power
spectrum by a Gaussian damping term $g(k) = \exp{(-k^2 \sigma_v^2)}$:
\begin{equation}
P_{\rm damped}(k) = g(k) \, P_L(k) + [1 - g(k)] \, P_{\rm ref}(k) ,
\label{eqpkdamp1}
\end{equation}
where the inclusion of the second term maintains the same small-scale
clustering amplitude.  The magnitude of the damping can be modelled
using perturbation theory (Crocce \& Scoccimarro 2008) as
\begin{equation}
\sigma_v^2 = \frac{1}{6 \pi^2} \int P_L(k) \, dk ,
\label{eqsigv}
\end{equation}
where $f = \om(z)^{0.55}$ is the growth rate of structure.  In our
fiducial cosmological model, $\omhh = 0.1361$, we find $\sigma_v =
4.5 \, h^{-1}$ Mpc.  We checked that this value was consistent with
the allowed range when $\sigma_v$ was varied as a free parameter and
fitted to the data.

Next, we incorporated the non-linear boost to the clustering power
using the fitting formula of Smith et al.\ (2003).  However, we
calculated the non-linear enhancement of power using the input
no-wiggles reference spectrum rather than the full linear model
including baryon oscillations:
\begin{equation}
P_{\rm damped,NL}(k) = \left( \frac{P_{\rm ref,NL}(k)}{P_{\rm ref}(k)}
\right) \times P_{\rm damped}(k) .
\label{eqpkdamp2}
\end{equation}
Equation \ref{eqpkdamp2} is then transformed into a correlation
function $\xi_{\rm damped,NL}(s)$ using Equation \ref{eqpktoxi}.

The final component of our model is a scale-dependent galaxy bias term
$B(s)$ relating the galaxy correlation function appearing in
Equation \ref{eqximod} to the non-linear matter correlation function:
\begin{equation}
\xi_{\rm fid,galaxy}(s) = B(s) \, \xi_{\rm damped,NL}(s) ,
\end{equation}
where we note that an overall constant normalization $b^2$ has already
been separated in Equation \ref{eqximod} so that $B(s) \rightarrow 1$
at large $s$.

We determined the form of $B(s)$ using halo catalogues extracted from
the GiggleZ dark matter simulation.  This $N$-body simulation has been
generated specifically in support of WiggleZ survey science, and
consists of $2160^3$ particles evolved in a $1 \, h^{-3}$ Gpc$^3$ box
using a WMAP5 cosmology (Komatsu et al.\ 2009).  We deduced $B(s)$
using the non-linear redshift-space halo correlation functions and
non-linear dark-matter correlation function of the simulation.  We
found that a satisfactory fitting formula for the scale-dependent bias
over the scales of interest is
\begin{equation}
B(s) = 1 + (s/s_0)^\gamma .
\label{eqscaledep}
\end{equation}
We performed this procedure for several contiguous subsets of
$250{,}000$ halos rank-ordered by their maximum circular velocity (a
robust proxy for halo mass).  The best-fitting parameters of Equation
\ref{eqscaledep} for the subset which best matches the large-scale
WiggleZ clustering amplitude are $s_0 = 0.32 \, h^{-1}$ Mpc, $\gamma =
-1.36$.  We note that the magnitude of the scale-dependent correction
from this term is $\sim 1\%$ for a scale $s \sim 10 \, h^{-1}$ Mpc,
which is far smaller than the $\sim 10\%$ magnitude of such effects
for more strongly-biased galaxy samples such as Luminous Red Galaxies
(Eisenstein et al.\ 2005).  This greatly reduces the potential for
systematic error due to a failure to model correctly scale-dependent
galaxy bias effects.

\begin{figure}
\begin{center}
\resizebox{\columnwidth}{!}{\includegraphics{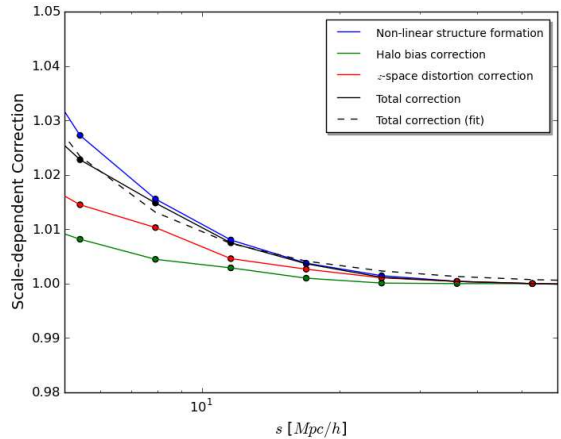}}
\end{center}
\caption{The scale-dependent correction to the non-linear real-space
  dark matter correlation function for haloes with maximum circular
  velocity $V_{\rm max} \approx 125$ km s$^{-1}$, which possess the
  same amplitude of large-scale clustering as WiggleZ galaxies.  The
  green line is the ratio of the real-space halo correlation function
  to the real-space non-linear dark matter correlation function.  The
  red line is the ratio of the redshift-space halo correlation
  function to the real-space halo correlation function.  The black
  line, the product of the red and green lines, is the scale-dependent
  bias correction $B(s)$ which we fitted with the model of Equation
  \ref{eqscaledep}, shown as the dashed black line.  The blue line is
  the ratio of the real-space non-linear to linear correlation
  function.}
\label{figscaledep}
\end{figure}

\subsection{Extraction of $D_V$}
\label{secxidv}

We fitted the galaxy correlation function template model described
above to the WiggleZ survey measurement, varying the matter density
$\omhh$, the scale distortion parameter $\alpha$ and the galaxy bias
$b^2$.  Our default fitting range was $10 < s < 180 \, h^{-1}$ Mpc
(following Eisenstein et al.\ 2005), where $10 \, h^{-1}$ is an
estimate of the minimum scale of validity for the quasi-linear theory
described in Section \ref{secxitemp}.  In the following, we assess the
sensitivity of the parameter constraints to the fitting range.

We minimized the $\chi^2$ statistic using the full data covariance
matrix, assuming that the probability of a model was proportional to
$\exp{(-\chi^2/2)}$.  The best-fitting parameters were $\omhh = 0.132
\pm 0.011$, $\alpha = 1.075 \pm 0.055$ and $b^2 = 1.21 \pm 0.11$,
where the errors in each parameter are produced by marginalizing over
the remaining two parameters.  The minimum value of $\chi^2$ is $14.9$
for 14 degrees of freedom (17 bins minus 3 fitted parameters),
indicating an acceptable fit to the data.  In Figure \ref{figxifit} we
compare the best-fitting correlation function model to the WiggleZ
data points.  The results of the parameter fits are summarized for
ease of reference in Table \ref{tabres}.

\begin{figure}
\begin{center}
\resizebox{\columnwidth}{!}{\rotatebox{270}{\includegraphics{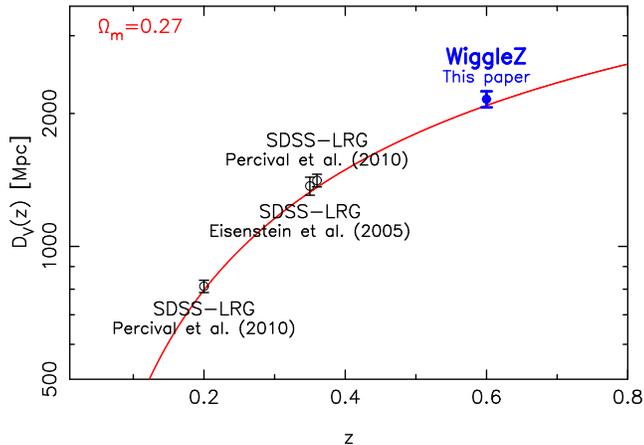}}}
\end{center}
\caption{Measurements of the distance-redshift relation using the BAO
  standard ruler from LRG samples (Eisenstein et al.\ 2005, Percival
  et al.\ 2010) and the current WiggleZ analysis.  The results are
  compared to a fiducial flat $\Lambda$CDM cosmological model with
  matter density $\om = 0.27$.}
\label{figdvcomp}
\end{figure}

Our measurement of the scale distortion parameter $\alpha$ may be
translated into a constraint on the distance scale $D_V = \alpha \,
D_{V,\rm fid} = 2234.9 \pm 115.2$ Mpc, corresponding to a $5.2\%$
measurement of the distance scale at $z=0.60$.  This accuracy is
comparable to that reported by Eisenstein et al.\ (2005) for the
analysis of the SDSS DR3 LRG sample at $z=0.35$.  Figure
\ref{figdvcomp} compares our measurement of the distance-redshift
relation with those from the LRG samples analyzed by Eisenstein et
al.\ (2005) and Percival et al.\ (2010).

\begin{figure}
\begin{center}
\resizebox{\columnwidth}{!}{\rotatebox{270}{\includegraphics{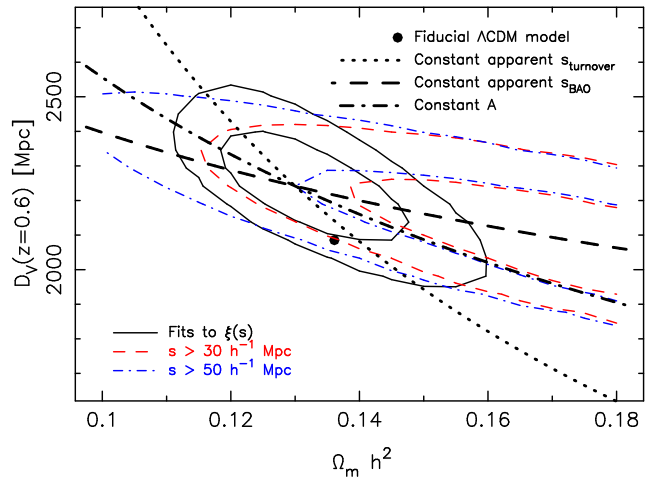}}}
\end{center}
\caption{Probability contours of the physical matter density $\omhh$
  and distance scale $D_V(z=0.6)$ obtained by fitting to the WiggleZ
  survey combined correlation function $\xi(s)$.  Results are compared
  for different ranges of fitted scales $s_{\rm min} < s < 180 \,
  h^{-1}$ Mpc.  The (black solid, red dashed, blue dot-dashed)
  contours correspond to fitting for $s_{\rm min} = (10, 30, 50) \,
  h^{-1}$ Mpc, respectively.  The heavy dashed and dotted lines are
  the degeneracy directions which are expected to result from fits
  involving respectively just the acoustic oscillations, and just the
  shape of a pure CDM power spectrum.  The heavy dash-dotted line
  represents a constant value of the acoustic ``$A$'' parameter
  introduced by Eisenstein et al.\ (2005), which is the parameter
  best-measured by our correlation function data.  The solid circle
  represents the location of our fiducial cosmological model.  The two
  contour levels in each case enclose regions containing $68\%$ and
  $95\%$ of the likelihood.}
\label{figprobxi}
\end{figure}

The 2D probability contours for the parameters $\omhh$ and
$D_V(z=0.6)$, marginalizing over $b^2$, are displayed in Figure
\ref{figprobxi}.  Following Eisenstein et al.\ (2005) we indicate
three degeneracy directions in this parameter space.  The first
direction (the dashed line in Figure \ref{figprobxi}) corresponds to a
constant measured acoustic peak separation, i.e.\ $r_s(z_d)/D_V(z=0.6)
= {\rm constant}$.  We used the fitting formula quoted in Percival et
al.\ (2010) to determine $r_s(z_d)$ as a function of $\omhh$ (given
our fiducial value of $\obhh = 0.0226$); we find that $r_s(z_d) =
152.6$ Mpc for our fiducial cosmological model.  The second degeneracy
direction (the dotted line in Figure \ref{figprobxi}) corresponds to a
constant measured shape of a Cold Dark Matter power spectrum,
i.e.\ $D_V(z=0.6) \times \omhh = {\rm constant}$.  In such models the
matter transfer function at recombination can be expressed as a
function of $q = k / \omhh$ (Bardeen et al.\ 1986).  Given that
changing $D_V$ corresponds to a scaling of $k \propto D_{V,{\rm
    fid}}/D_V$, we recover that the measured power spectrum shape
depends on $D_V \omhh$.  The principle degeneracy axis of our
measurement lies between these two curves, suggesting that both the
correlation function shape and acoustic peak information are driving
our measurement of $D_V$.  The third degeneracy direction we plot (the
dash-dotted line in Figure \ref{figprobxi}), which matches our
measurement, corresponds to a constant value of the acoustic parameter
$A(z) \equiv D_V(z) \sqrt{\om H_0^2}/cz$ introduced by Eisenstein et
al.\ (2005).  We present our fits for this parameter in Section
\ref{secdist}.

In Figure \ref{figprobxi} we also show probability contours resulting
from fits to a restricted range of separations $s > 30$ and $50 \,
h^{-1}$ Mpc.  In both cases the contours become significantly more
extended and the long axis shifts into alignment with the case of the
acoustic peak alone driving the fits.  The restricted fitting range no
longer enables us to perform an accurate determination of the value of
$\omhh$ from the shape of the clustering pattern alone.

\subsection{Significance of the acoustic peak detection}

In order to assess the importance of the baryon acoustic peak in
constraining this model, we repeated the parameter fit replacing the
model correlation function with one generated using a ``no-wiggles''
reference power spectrum $P_{\rm ref}(k)$, which possesses the same
amplitude and overall shape as the original matter power spectrum but
lacks the baryon oscillation features (i.e., we replaced $P_L(k)$ with
$P_{\rm ref}(k)$ in Equation \ref{eqpkdamp1}).  The minimum value
obtained for the $\chi^2$ statistic in this case was $25.0$,
indicating that the model containing baryon oscillations was favoured
by $\Delta \chi^2 = 10.1$.  This corresponds to a detection of the
acoustic peak with a statistical significance of $3.2$-$\sigma$.
Furthermore, the value and error obtained for the scale distortion
parameter in the no-wiggles model was $\alpha = 0.80 \pm 0.17$,
representing a degradation of the error in $\alpha$ by a factor of
three.  This also suggests that the acoustic peak is important for
establishing the distance constraints from our measurement.

As an alternative approach for assessing the significance of the
acoustic peak, we changed the fiducial baryon density to $\ob = 0$ and
repeated the parameter fit.  The minimum value obtained for the
$\chi^2$ statistic was now $22.7$ and the value and marginalized error
determined for the scale distortion parameter was $\alpha = 0.80 \pm
0.12$, re-affirming the significance of our detection of the baryon
wiggles.

If we restrict the correlation function fits to the range $50 < s <
130 \, h^{-1}$ Mpc, further reducing the influence of the overall
shape of the clustering pattern on the goodness-of-fit, we find that
our fiducial model has a minimum $\chi^2 = 5.9$ (for 5 degrees of
freedom) and the ``no-wiggles'' reference spectrum produces a minimum
$\chi^2 = 13.1$.  Even for this restricted range of scales, the model
containing baryon oscillations was therefore favoured by $\Delta
\chi^2 = 7.2$.

\subsection{Sensitivity to the clustering model}

In this Section we investigate the systematic dependence of our
measurement of $D_V(z=0.6)$ on the model used to describe the
quasi-linear correlation function.  We considered five modelling
approaches proposed in the literature:

\begin{itemize}

\item {\it Model 1:} Our fiducial model described in Section
  \ref{secxitemp} following Eisenstein et al.\ (2005), in which the
  quasi-linear damping of the acoustic peak was modelled by an
  exponential factor $g(k) = \exp{(-k^2 \sigma_v^2)}$, $\sigma_v$ is
  determined from linear theory via Equation \ref{eqsigv}, and the
  small-scale power was restored by adding a term $[1-g(k)]$ multiplied
  by the wiggle-free reference spectrum (Equation \ref{eqpkdamp1}).

\item {\it Model 2:} No quasi-linear damping of the acoustic peak was
  applied, i.e.\ $\sigma_v = 0$.

\item {\it Model 3:} The term restoring the small-scale power,
  $[1-g(k)] P_{\rm ref}(k)$ in Equation \ref{eqpkdamp1}, was omitted.

\item {\it Model 4:} $P_{\rm damped}(k)$ in Equation \ref{eqpkdamp1}
  was generated using Equation 14 of Eisenstein, Seo \& White (2006),
  which implements different damping coefficients in the tangential
  and radial directions.

\item {\it Model 5:} The quasi-linear matter correlation function was
  generated using Equation 10 of Sanchez et al.\ (2009), following
  Crocce \& Scoccimarro (2008), which includes the additional
  contribution of a ``mode-coupling'' term.  We set the coefficient
  $A_{\rm MC} = 1$ in this equation (rather than introduce an
  additional free parameter).

\end{itemize}

\begin{figure}
\begin{center}
\resizebox{\columnwidth}{!}{\rotatebox{270}{\includegraphics{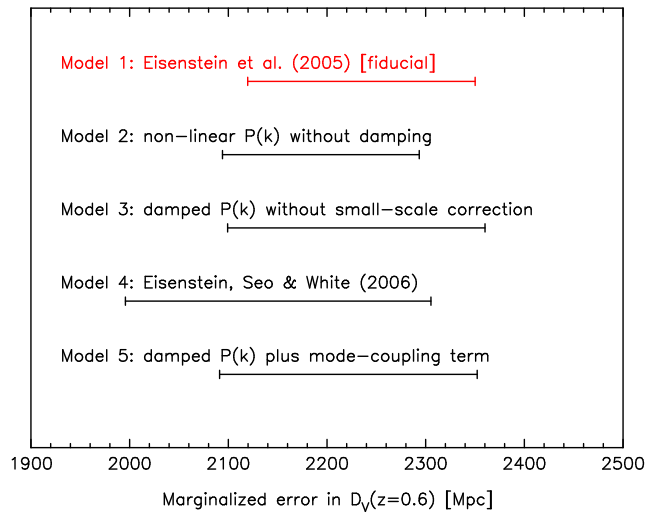}}}
\end{center}
\caption{Measurements of $D_V(z=0.6)$ from the galaxy correlation
  function, marginalizing over $\omhh$ and $b^2$, comparing five
  different models for the quasi-linear correlation function as
  detailed in the text.  The measurements are consistent, suggesting
  that systematic modelling errors are not significantly affecting our
  results.}
\label{figdvsys}
\end{figure}

Figure \ref{figdvsys} compares the measurements of $D_V(z=0.6)$ from
the correlation function data, marginalized over $\omhh$ and $b^2$,
assuming each of these models.  The agreement amongst the best-fitting
measurements is excellent, and the minimum $\chi^2$ statistics imply a
good fit to the data in each case.  We conclude that systematic errors
associated with modelling the correlation function are not
significantly affecting our results.  The error in the distance
measurement is determined by the amount of damping of the acoustic
peak, which controls the precision with which the standard ruler may
be applied.  The lowest distance error is produced by Model 2 which
neglects damping; the greatest distance error is associated with Model
4, in which the damping is enhanced along the line-of-sight (see
Equation 13 in Eisenstein, Seo \& White 2006).

\begin{table*}
\begin{center}
\caption{Results of fitting a three-parameter model $(\omhh, \alpha,
  b^2)$ to WiggleZ measurements of four different clustering
  statistics for various ranges of scales.  The top four entries,
  above the horizontal line, correspond to our fiducial choices of
  fitting range for each statistic.  The fitted scales $\alpha$ are
  converted into measurements of $D_V$ and two BAO distilled
  parameters, $A$ and $r_s(z_d)/D_V$, which are introduced in Section
  \ref{secdist}.  The final column lists the measured value of $D_V$
  when the parameter $\omhh$ is left fixed at its fiducial value and
  only the bias $b^2$ is marginalized.  We recommend using $A(z=0.6)$
  as measured by the correlation function $\xi(s)$ for the scale range
  $10 < s < 180 \, h^{-1}$ Mpc, highlighted in bold, as the most
  appropriate WiggleZ measurement for deriving BAO constraints on
  cosmological parameters.}
\label{tabres}
\begin{tabular}{ccccccc}
\hline
Statistic & Scale range & $\omhh$ & $D_V(z=0.6)$ & $A(z=0.6)$ & $r_s(z_d)/D_V(z=0.6)$ & $D_V(z=0.6)$ \\
& & & [Mpc] & & & fixing $\omhh$ \\
\hline
$\xi(s)$
& $10 < s < 180 \, h^{-1}$ Mpc
& $0.132 \pm 0.011$
& $2234.9 \pm 115.2$
& ${\mathbf 0.452 \pm 0.018}$
& $0.0692 \pm 0.0033$
& $2216.5 \pm 78.9$ \\
$P(k)$ [full]
& $0.02 < k < 0.2 \, h$ Mpc$^{-1}$
& $0.134 \pm 0.008$
& $2160.7 \pm 132.3$
& $0.440 \pm 0.020$
& $0.0711 \pm 0.0038$
& $2141.0 \pm 97.5$ \\
$P(k)$ [wiggles]
& $0.02 < k < 0.2 \, h$ Mpc$^{-1}$
& $0.163 \pm 0.017$
& $2135.4 \pm 156.7$
& $0.461 \pm 0.030$
& $0.0699 \pm 0.0045$
& $2197.2 \pm 119.1$ \\
$w_0(r)$
& $10 < r < 180 \, h^{-1}$ Mpc
& $0.130 \pm 0.011$
& $2279.2 \pm 142.4$
& $0.456 \pm 0.021$
& $0.0680 \pm 0.0037$
& $2238.2 \pm 104.6$ \\
\hline
$\xi(s)$
& $30 < s < 180 \, h^{-1}$ Mpc
& $0.166 \pm 0.014$
& $2127.7 \pm 127.9$
& $0.475 \pm 0.025$
& $0.0689 \pm 0.0031$
& $2246.8 \pm 102.6$ \\
$\xi(s)$
& $50 < s < 180 \, h^{-1}$ Mpc
& $0.164 \pm 0.016$
& $2129.2 \pm 140.8$
& $0.474 \pm 0.025$
& $0.0690 \pm 0.0031$
& $2240.1 \pm 104.7$ \\
$P(k)$ [full]
& $0.02 < k < 0.1 \, h$ Mpc$^{-1}$
& $0.150 \pm 0.020$
& $2044.7 \pm 253.0$
& $0.441 \pm 0.034$
& $0.0733 \pm 0.0073$
& $2218.1 \pm 128.4$ \\
$P(k)$ [full]
& $0.02 < k < 0.3 \, h$ Mpc$^{-1}$
& $0.137 \pm 0.007$
& $2132.1 \pm 109.2$
& $0.441 \pm 0.017$
& $0.0716 \pm 0.0033$
& $2148.9 \pm 79.9$ \\
$P(k)$ [wiggles]
& $0.02 < k < 0.1 \, h$ Mpc$^{-1}$
& $0.160 \pm 0.020$
& $2240.7 \pm 235.8$
& $0.466 \pm 0.034$
& $0.0678 \pm 0.0070$
& $2277.9 \pm 187.5$ \\
$P(k)$ [wiggles]
& $0.02 < k < 0.3 \, h$ Mpc$^{-1}$
& $0.161 \pm 0.019$
& $2114.5 \pm 132.4$
& $0.455 \pm 0.026$
& $0.0706 \pm 0.0037$
& $2171.4 \pm 98.0$ \\
$w_0(r)$
& $30 < r < 180 \, h^{-1}$ Mpc
& $0.127 \pm 0.018$
& $2288.8 \pm 157.3$
& $0.455 \pm 0.027$
& $0.0681 \pm 0.0037$
& $2251.6 \pm 111.7$ \\
$w_0(r)$
& $50 < r < 180 \, h^{-1}$ Mpc
& $0.164 \pm 0.016$
& $2190.0 \pm 146.2$
& $0.466 \pm 0.023$
& $0.0673 \pm 0.0036$
& $2282.1 \pm 109.8$ \\
\hline
\end{tabular}
\end{center}
\end{table*}

\section{Power spectrum}
\label{secpk}

\subsection{Measurements and covariance matrix}
\label{secpkmeas}

The power spectrum is a second commonly-used method for quantifying
the galaxy clustering pattern, which is complementary to the
correlation function.  It is calculated using a Fourier decomposition
of the density field in which (contrary to the correlation function)
the maximal signal-to-noise is achieved on large, linear or
quasi-linear scales (at low wavenumbers) and the measurement of
small-scale power (at high wavenumbers) is limited by shot noise.
However, also in contrast to the correlation function, small-scale
effects such as shot noise influence the measured power at all
wavenumbers, and the baryon oscillation signature appears as a series
of decaying harmonic peaks and troughs at different wavenumbers.  In
aesthetic terms this diffusion of the baryon oscillation signal is
disadvantageous.

We estimated the galaxy power spectrum for each separate WiggleZ
survey region using the direct Fourier methods introduced by Feldman,
Kaiser \& Peacock (1994; FKP).  Our methodology is fully described in
Section 3.1 of Blake et al.\ (2010); we give a brief summary here.
Firstly we map the angle-redshift survey cone into a cuboid of
co-moving co-ordinates using a fiducial flat $\Lambda$CDM cosmological
model with matter density $\om = 0.27$.  We gridded the catalogue in
cells using nearest grid point assignment ensuring that the Nyquist
frequencies in each direction were much higher than the Fourier
wavenumbers of interest (we corrected the power spectrum measurement
for the small bias introduced by this gridding).  We then applied a
Fast Fourier transform to the grid, optimally weighting each pixel by
$1/(1 + n P_0)$, where $n$ is the galaxy number density in the pixel
(determined using the selection function) and $P_0 = 5000 \, h^{-3}$
Mpc$^3$ is a characteristic power spectrum amplitude.  The Fast
Fourier transform of the selection function is then used to construct
the final power spectrum estimator using Equation 13 in Blake et
al.\ (2010).  The measurement is corrected for the effect of redshift
blunders using Monte Carlo survey simulations as described in Section
3.2 of Blake et al.\ (2010).  We measured each power spectrum in
wavenumber bins of width $0.01 \, h$ Mpc$^{-1}$ between $k=0$ and $0.3
\, h$ Mpc$^{-1}$, and determined the covariance matrix of the
measurement in these bins by implementing the sums in Fourier space
described by FKP (see Blake et al.\ 2010 equations 20-22).  The FKP
errors agree with those obtained from lognormal realizations within
$10\%$ at all scales.

In order to detect and fit for the baryon oscillation signature in the
WiggleZ galaxy power spectrum, we need to stack together the
measurements in the individual survey regions and redshift slices.
This requires care because each sub-region possesses a different
selection function, and therefore each power spectrum measurement
corresponds to a different convolution of the underlying power
spectrum model.  Furthermore the non-linear component of the
underlying model varies with redshift, due to non-linear evolution of
the density and velocity power spectra.  Hence the observed power
spectrum in general has a systematically-different slope in each
sub-region, which implies that the baryon oscillation peaks lie at
slightly different wavenumbers.  If we stacked together the raw
measurements, there would be a significant washing-out of the acoustic
peak structure.

Therefore, before combining the measurements, we made a correction to
the shape of the various power spectra to bring them into alignment.
We wish to avoid spuriously enhancing the oscillatory features when
making this correction.  Our starting point is therefore a fiducial
power spectrum model generated from the Eisenstein \& Hu (1998)
``no-wiggles'' reference linear power spectrum, which defines the
fiducial slope to which we correct each measurement.  Firstly, we
modified this reference function into a redshift-space non-linear
power spectrum, using an empirical redshift-space distortion model
fitted to the two-dimensional power spectrum split into tangential and
radial bins (see Blake et al.\ 2011a).  The redshift-space distortion
is modelled by a coherent-flow parameter $\beta$ and a pairwise
velocity dispersion parameter $\sigma_v$, which were fitted
independently in each of the redshift slices.  We convolved this
redshift-space non-linear reference power spectrum with the selection
function in each sub-region, and our correction factor for the
measured power spectrum is then the ratio of this convolved function
to the original real-space linear reference power spectrum.  After
applying this correction to the data and covariance matrix we combined
the resulting power spectra using inverse-variance weighting.

\begin{figure}
\begin{center}
\resizebox{\columnwidth}{!}{\rotatebox{270}{\includegraphics{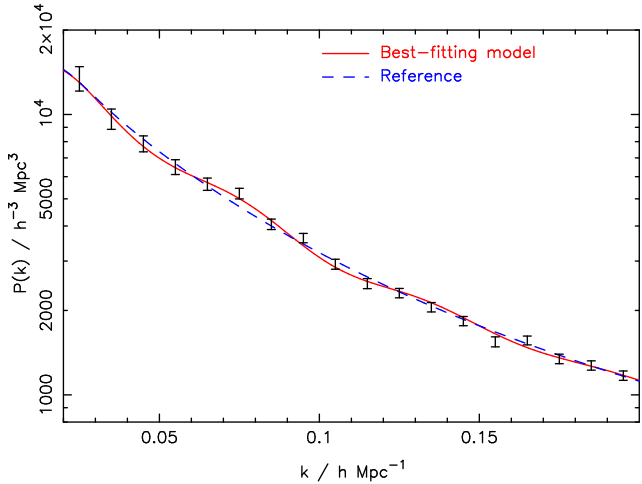}}}
\end{center}
\caption{The power spectrum obtained by stacking measurements in
  different WiggleZ survey regions using the method described in
  Section \ref{secpkmeas}.  The best-fitting power spectrum model
  (varying $\omhh$, $\alpha$ and $b^2$) is overplotted as the solid
  line.  We also show the corresponding ``no-wiggles'' reference model
  as the dashed line, constructed from a power spectrum with the same
  clustering amplitude but lacking baryon acoustic oscillations.}
\label{figpkfit}
\end{figure}

\begin{figure}
\begin{center}
\resizebox{\columnwidth}{!}{\rotatebox{270}{\includegraphics{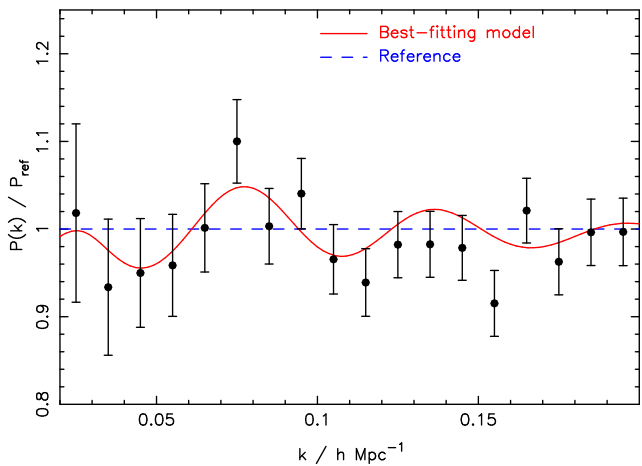}}}
\end{center}
\caption{The combined WiggleZ survey power spectrum of Figure
  \ref{figpkfit} divided by the smooth reference spectrum to reveal
  the signature of baryon oscillations more clearly.  We detect the
  first harmonic peak in Fourier space.}
\label{figpkreffit}
\end{figure}

\begin{figure}
\begin{center}
\resizebox{\columnwidth}{!}{\includegraphics{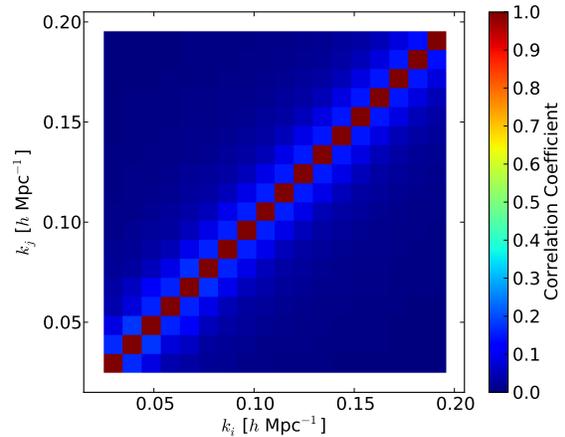}}
\end{center}
\caption{The amplitude of the cross-correlation $C_{ij}/\sqrt{C_{ii}
    C_{jj}}$ of the covariance matrix $C_{ij}$ for the power spectrum
  measurement, determined using the FKP estimator.  The amplitude of
  the off-diagonal elements of the covariance matrix is very low.}
\label{figpkcov}
\end{figure}

Figures \ref{figpkfit} and \ref{figpkreffit} respectively display the
combined power spectrum data, and that data divided through by the
combined no-wiggles reference spectrum in order to reveal any
signature of acoustic oscillations more clearly.  We note that there
is a significant enhancement of power at the position of the first
harmonic, $k \approx 0.075 \, h$ Mpc$^{-1}$.  The other harmonics are
not clearly detected with the current dataset, although the model is
nevertheless a good statistical fit.  Figure \ref{figpkcov} displays
the final power spectrum covariance matrix, resulting from combining
the different WiggleZ survey regions, in the form of a correlation
matrix $C_{ij}/\sqrt{C_{ii} C_{jj}}$.  We note that there is very
little correlation between separate $0.01 \, h$ Mpc$^{-1}$ power
spectrum bins.

We note that our method for combining power spectrum measurements in
different sub-regions only corrects for the convolution effect of the
window function on the overall power spectrum shape, and does not undo
the smoothing of the BAO signature in each window.  We therefore
expect the resulting BAO detection in the combined power spectrum may
have somewhat lower significance than that in the combined correlation
function.

\subsection{Extraction of $D_V$}

We investigated two separate methods for fitting the scale distortion
parameter to the power spectrum data.  Our first approach used the
whole shape of the power spectrum including any baryonic signature.
We generated a template model non-linear power spectrum $P_{\rm
  fid}(k)$ parameterized by $\omhh$, which we took as Equation
\ref{eqpkdamp2} in Section \ref{secxitemp}, and fitted the model
\begin{equation}
P_{\rm mod}(k) = b^2 P_{\rm fid}(k/\alpha) ,
\end{equation}
where $\alpha$ now appears in the denominator (as opposed to the
numerator of Equation \ref{eqximod}) due to the switch from real space
to Fourier space.  As in the case of the correlation function, the
probability distribution of $\alpha$, after marginalizing over $\omhh$
and $b^2$, can be connected to the measurement of $D_V(z_{\rm eff})$.
We determined the effective redshift of the power spectrum estimate by
weighting each pixel in the selection function by its contribution to
the power spectrum error:
\begin{equation}
  z_{\rm eff} = \sum_{\vec{x}} z \, \left( \frac{n_g(\vec{x})
      P_g}{1 + n_g(\vec{x}) P_g} \right)^2 ,
\end{equation}
where $n_g(\vec{x})$ is the galaxy number density in each grid cell
$\vec{x}$ and $P_g$ is the characteristic galaxy power spectrum
amplitude, which we evaluated at a scale $k = 0.1 \, h$ Mpc$^{-1}$.
We obtained an effective redshift $z_{\rm eff} = 0.583$.  In order to
enable comparison with the correlation function fits we applied the
best-fitting value of $\alpha$ at $z=0.6$.

Our second approach to fitting the power spectrum measurement used
only the information contained in the baryon oscillations.  We divided
the combined WiggleZ power spectrum data by the corresponding combined
no-wiggles reference spectrum, and when fitting models we divided each
trial power spectrum by its corresponding reference spectrum prior to
evaluating the $\chi^2$ statistic.

\begin{figure}
\begin{center}
\resizebox{\columnwidth}{!}{\rotatebox{270}{\includegraphics{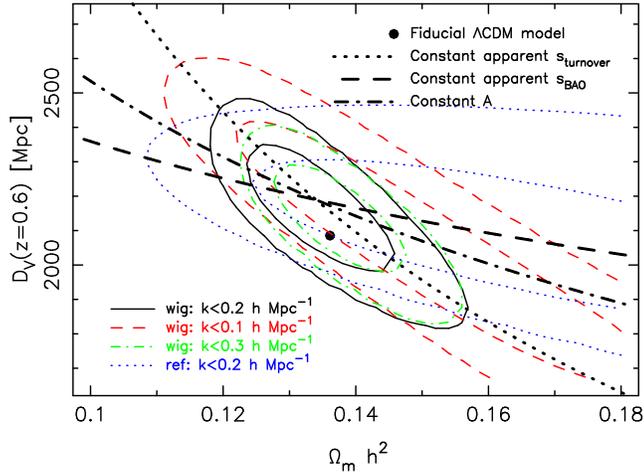}}}
\end{center}
\caption{Probability contours of the physical matter density $\omhh$
  and distance scale $D_V(z=0.6)$ obtained by fitting to the WiggleZ
  survey combined power spectrum.  Results are compared for different
  ranges of fitted scales $0.02 < k < k_{\rm max}$ and methods.  The
  (red dashed, black solid, green dot-dashed) contours correspond to
  fits of the full power spectrum model for $k_{\rm max} = (0.1, 0.2,
  0.3) \, h$ Mpc$^{-1}$, respectively.  The blue dotted contours
  result from fitting to the power spectrum divided by a smooth
  no-wiggles reference spectrum (with $k_{\rm max} = 0.2 \, h$
  Mpc$^{-1}$).  Degeneracy directions and likelihood contour levels
  are plotted as in Figure \ref{figprobxi}.}
\label{figprobpk}
\end{figure}

We restricted our fits to Fourier wavescales $0.02 < k < 0.2 \, h$
Mpc$^{-1}$, where the upper limit is an estimate of the range of
reliability of the quasi-linear power spectrum modelling.  We
investigate below the sensitivity of the best-fitting parameters to
the fitting range.  For the first method, fitting to the full power
spectrum shape, the best-fitting parameters and $68\%$ confidence
ranges were $\omhh = 0.134 \pm 0.008$ and $\alpha = 1.050 \pm 0.064$,
where the errors in each parameter are produced by marginalizing over
the remaining two parameters.  The minimum value of $\chi^2$ was
$12.4$ for 15 degrees of freedom (18 bins minus 3 fitted parameters),
indicating an acceptable fit to the data.  We can convert the
constraint on the scale distortion parameter into a measured distance
$D_V(z=0.6) = 2160.7 \pm 132.3$ Mpc.  The 2D probability distribution
of $\omhh$ and $D_V(z=0.6)$, marginalizing over $b^2$, is displayed as
the solid contours in Figure \ref{figprobpk}.  In this Figure we
reproduce the same degeneracy lines discussed in Section
\ref{secxidv}, which are expected to result from fits involving just
the acoustic oscillations and just the shape of a pure CDM power
spectrum.  We note that the long axis of our probability contours is
oriented close to the latter line, indicating that the acoustic peak
is not exerting a strong influence on fits to the full WiggleZ power
spectrum shape.  Comparison of Figure \ref{figprobpk} with Figure
\ref{figprobxi} shows that fits to the WiggleZ galaxy correlation
function are currently more influenced by the BAOs than the power
spectrum.  This is attributable to the signal being stacked at a
single scale in the correlation function, in this case of a moderate
BAO detection.

For the second method, fitting to just the baryon oscillations, the
best-fitting parameters and $68\%$ confidence ranges were $\omhh =
0.163 \pm 0.017$ and $\alpha = 1.000 \pm 0.073$.  Inspection of the 2D
probability contours of $\omhh$ and $\alpha$, which are shown as the
dotted contours in Figure \ref{figprobpk}, indicates that a
significant degeneracy has opened up parallel to the line of constant
apparent BAO scale (as expected).  Increasing $\omhh$ decreases the
standard ruler scale, but the positions of the acoustic peaks may be
brought back into line with the data by applying a lower scale
distortion parameter $\alpha$.  Low values of $\omhh$ are ruled out
because the resulting amplitude of baryon oscillations is too high
(given that $\obhh$ is fixed).  We also plot in Figure \ref{figprobpk}
the probability contours resulting from fitting different ranges of
Fourier scales $k < 0.1 \, h$ Mpc$^{-1}$ and $k < 0.3 \, h$
Mpc$^{-1}$.  The $68\%$ confidence regions generated for these
different cases overlap.

We assessed the significance with which acoustic features are detected
in the power spectrum using a method similar to our treatment of the
correlation function in Section \ref{secxi}.  We repeated the
parameter fit for $(\omhh, \alpha, b^2)$ using the ``no-wiggles''
reference power spectrum in place of the full model power spectrum.
The minimum value obtained for the $\chi^2$ statistic in this case was
$15.8$, indicating that the model containing baryon oscillations was
favoured by only $\Delta \chi^2 = 3.3$.  This is consistent with the
direction of the long axis of the probability contours in Figure
\ref{figprobpk}, which suggests that the baryon oscillations are not
driving the fits to the full power spectrum shape.

\section{Band-filtered correlation function}
\label{secxu}

\subsection{Measurements and covariance matrix}

Xu et al.\ (2010) introduced a new statistic for the measurement of
the acoustic peak in configuration space, which they describe as an
advantageous approach for band-filtering the information.  They proposed
estimating the quantity
\begin{equation}
  w_0(r) = 4 \pi \int_0^r \frac{ds}{r} \, \left( \frac{s}{r} \right)^2
  \xi(s) \, W \left( \frac{s}{r} \right) ,
\label{eqxitoxu}
\end{equation}
where $\xi(s)$ is the 2-point correlation function as a function of
separation $s$ and
\begin{eqnarray}
  W(x) &=& (2x)^2 (1-x)^2 \left( \frac{1}{2}-x \right) \hspace{5mm} 0 < x < 1 \\
  &=& 0 \hspace{5mm} {\rm otherwise}
\end{eqnarray}
in terms of $x = (s/r)^3$.  This filter localizes the acoustic
information in a single feature at the acoustic scale in a similar
manner to the correlation function, which is beneficial for securing a
robust, model-independent detection.  However, the form of the filter
function $W(x)$ advantageously reduces the sensitivity to small-scale
power (which is difficult to model due to non-linear effects) and
large-scale power (which is difficult to measure because it is subject
to uncertainties regarding the mean density of the sample), combining
the respective advantages of the correlation function and power
spectrum approaches.

Xu et al.\ (2010) proposed that $w_0(r)$ should be estimated as a
weighted sum over galaxy pairs $i$
\begin{equation}
  w_0(r) = DD_{\rm filtered}(r) = \frac{2}{N_D \, n_D}
  \sum_{i=1}^{N_{\rm pairs}} \frac{W(s_i/r)}{\phi(s_i,\mu_i)} ,
\label{eqxuest}
\end{equation}
where $s_i$ is the separation of pair $i$, $\mu_i$ is the cosine of
the angle of the separation vector to the line-of-sight, $N_D$ is the
number of data galaxies, and $n_D$ is the average galaxy density which
we simply define as $N_D/V$ where $V$ is the volume of a cuboid
enclosing the survey cone.  The function $\phi(s,\mu)$ describes the
edge effects due to the survey boundaries and is normalized so that
the number of random pairs in a bin $(s \rightarrow s + ds, \mu
\rightarrow \mu + d\mu)$ is
\begin{equation}
RR(s,\mu) = 2 \pi n_D N_D s^2 \phi(s,\mu) \, ds \, d\mu ,
\end{equation}
where $\phi = 1$ for a uniform, infinite survey.  We determined the
function $\phi(s,\mu)$ used in Equation \ref{eqxuest} by binning the
pair counts $RR(s,\mu)$ for many random sets in fine bins of $s$ and
$\mu$.  We then fitted a parameterized model
\begin{equation}
\phi(s,\mu) = \sum_{n=0}^3 a_n(s) \, \mu^{2n}
\end{equation}
to the result in bins of $s$, and used the coefficients $a_n(s)$ to
generate the value of $\phi$ for each galaxy pair.

We note that our Equation \ref{eqxuest} contains an extra factor of 2
compared to Equation 12 of Xu et al.\ (2010) because we define the
quantity as a sum over unique pairs, rather than all pairs.  We also
propose to modify the estimator to introduce a ``DR'' term by analogy
with the correlation function estimator of Equation \ref{eqxiest}, in
order to correct for the distribution of data galaxies with respect to
the boundaries of the sample:
\begin{equation}
w_0(r) = DD_{\rm filtered}(r) - DR_{\rm filtered}(r) ,
\label{eqxuestmod}
\end{equation}
where $DR_{\rm filtered}(r)$ is estimated using Equation
\ref{eqxuest}, but summing over data-random pairs and excluding the
initial factor of 2.  We used our lognormal realizations to determine
that this modified estimator of Equation \ref{eqxuestmod} produces a
result with lower bias and variance compared to Equation
\ref{eqxuest}.  We used Equation \ref{eqxuestmod} to measure the band
filtered correlation function of each WiggleZ region for 17 values of
$r$ spaced by $10 \, h^{-1}$ Mpc between $15$ and $175 \, h^{-1}$ Mpc.

We determined the covariance matrix $C_{ij}$ of our estimator using
the ensemble of lognormal realizations for each survey region.  We
note that for our dataset the amplitude of the diagonal errors
$\sqrt{C_{ii}}$ determined by lognormal realizations is typically
$\sim 5$ times greater than jack-knife errors and $\sim 3$ times
higher than obtained by evaluating Equation 13 in Xu et al.\ (2010)
which estimates the covariance matrix in the Gaussian limit.  Given
the likely drawbacks of jack-knife errors (the lack of independence of
the jack-knife regions on large scales) and Gaussian errors (which
fail to incorporate the survey selection function), the lognormal
errors should provide by far the best estimate of the covariance
matrix for this measurement.

\begin{figure}
\begin{center}
\resizebox{\columnwidth}{!}{\rotatebox{270}{\includegraphics{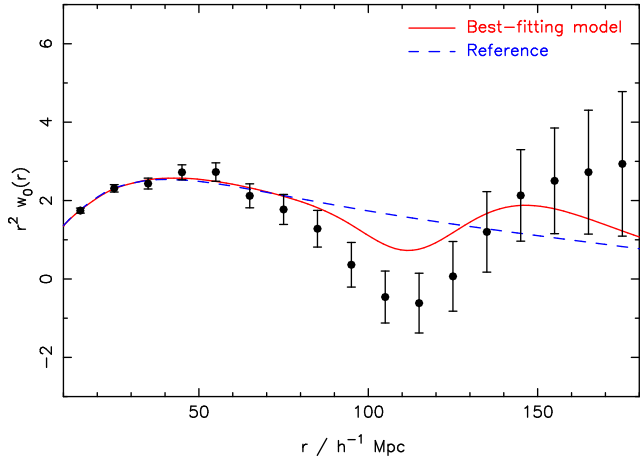}}}
\end{center}
\caption{The band-filtered correlation function $w_0(r)$ for the
  combined WiggleZ survey regions, plotted in the combination $r^2
  w_0(r)$.  The best-fitting clustering model (varying $\omhh$,
  $\alpha$ and $b^2$) is overplotted as the solid line.  We also show
  the corresponding ``no-wiggles'' reference model, constructed from a
  power spectrum with the same clustering amplitude but lacking baryon
  acoustic oscillations.  We note that the high covariance of the data
  points for this estimator implies that (despite appearances) the
  solid line is a good statistical fit to the data.}
\label{figxufit}
\end{figure}

\begin{figure}
\begin{center}
\resizebox{\columnwidth}{!}{\includegraphics{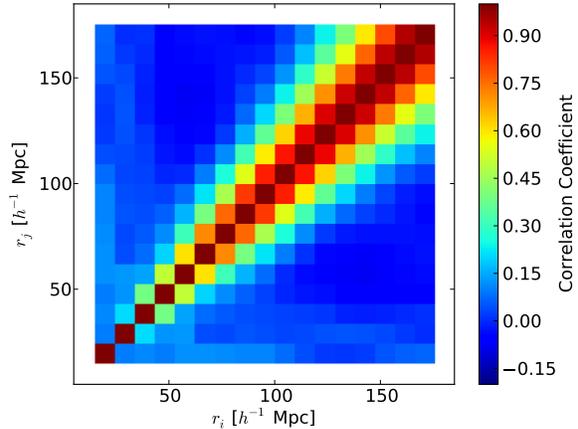}}
\end{center}
\caption{The amplitude of the cross-correlation $C_{ij}/\sqrt{C_{ii}
    C_{jj}}$ of the covariance matrix $C_{ij}$ for the band-filtered
  correlation function measurement plotted in Figure \ref{figxufit},
  determined using lognormal realizations.}
\label{figxucov}
\end{figure}

We constructed the final measurement of the band-filtered correlation
function by stacking the individual measurements in different survey
regions with inverse-variance weighting.  Figure \ref{figxufit}
displays our measurement.  We detect clear evidence of the expected
dip in $w_0(r)$ at the acoustic scale.  Figure \ref{figxucov} displays
the final covariance matrix of the band-filtered correlation function
resulting from combining the different WiggleZ survey regions in the
form of a correlation matrix $C_{ij}/\sqrt{C_{ii} C_{jj}}$.  We note
that the nature of the $w_0(r)$ estimator, which depends on the
correlation function at all scales $s < r$, implies that the data
points in different bins of $r$ are highly correlated, and the
correlation coefficient increases with $r$.  At the acoustic scale,
neighbouring $10 \, h^{-1}$ Mpc bins are correlated at the $\sim 85\%$
level and bins spaced by $20 \, h^{-1}$ Mpc are correlated at a level
of $\sim 55\%$.

\subsection{Extraction of $D_V$}

We determined the acoustic scale from the band-filtered correlation
function by constructing a template function $w_{0,{\rm
    fid,galaxy}}(r)$ in the same style as Section \ref{secxitemp}
and then fitting the model
\begin{equation}
w_{0,{\rm mod}}(r) = b^2 \, w_{0,{\rm fid,galaxy}}(\alpha \, r) .
\end{equation}
We determined the function $w_{0,{\rm fid,galaxy}}(r)$ by applying
the transformation of Equation \ref{eqxitoxu} to the template galaxy
correlation function $\xi_{\rm fid,galaxy}(s)$ defined in Section
\ref{secxitemp}, as a function of $\omhh$.

The best-fitting parameters to the band-filtered correlation function
are $\omhh = 0.130 \pm 0.011$, $\alpha = 1.100 \pm 0.069$ and $b^2 =
1.32 \pm 0.13$, where the errors in each parameter are produced by
marginalizing over the remaining two parameters.  The minimum value of
$\chi^2$ is $10.5$ for 14 degrees of freedom (17 bins minus 3 fitted
parameters), indicating an acceptable fit to the data.  Our
measurement of the distortion parameter may be translated into a
constraint on the distance scale $D_V(z=0.6) = \alpha D_{V,\rm fid} =
2279.2 \pm 142.4$ Mpc, corresponding to a $6.2\%$ measurement of the
distance scale at $z=0.6$.  Probability contours of $D_V(z=0.6)$ and
$\omhh$ are overplotted in Figure \ref{figprobdv}.

In Figure \ref{figxufit} we compare the best fitting band-filtered
correlation function model to the WiggleZ data points (noting that the
strong covariance between the data gives the mis-leading impression of
a poor fit).  We overplot a second model which corresponds to our
best-fit parameters but for which the no-wiggle reference power
spectrum has been used in place of the full power spectrum.  If we fit
this no-wiggles model varying $\omhh$, $\alpha$ and $b^2$ we find a
minimum value of $\chi^2 = 19.0$, implying that the model containing
acoustic features is favoured by $\Delta \chi^2 = 8.5$.

\section{Comparison of clustering statistics}
\label{seccomp}

\begin{figure}
\begin{center}
\resizebox{\columnwidth}{!}{\rotatebox{270}{\includegraphics{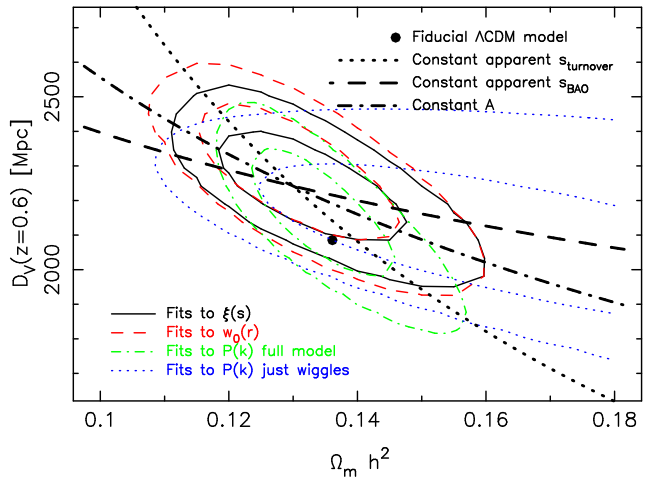}}}
\end{center}
\caption{A comparison of the probability contours of $\omhh$ and
  $D_V(z=0.6)$ resulting from fitting different clustering statistics
  measured from the WiggleZ survey: the correlation function for scale
  range $10 < s < 180 \, h^{-1}$ Mpc (the solid black contours), the
  band-filtered correlation function for scale range $10 < r < 180 \,
  h^{-1}$ Mpc (the dashed red contours), the full power spectrum shape
  for scale range $0.02 < k < 0.2 \, h$ Mpc$^{-1}$ (the dot-dashed
  green contours), and the power spectrum divided by a ``no-wiggles''
  reference spectrum (the dotted blue contours).  Degeneracy
  directions and likelihood contour levels are plotted as in Figure
  \ref{figprobxi}.}
\label{figprobdv}
\end{figure}

The distance-scale measurements of $D_V(z=0.6)$ using the four
different clustering statistics applied in this paper are compared in
Figure \ref{figprobdv}.  All four statistics give broadly consistent
results for the measurement of $D_V$ and $\omhh$, with significant
overlap between the respective $68\%$ confidence regions in this
parameter space.  This agreement suggests that systematic measurement
errors in these statistics are not currently dominating the WiggleZ
BAO fits.

There are some differences in detail between the results derived from
the four statistics.  Fits to the galaxy power spectrum are currently
dominated by the power spectrum shape rather than the BAOs, such that
the degeneracy direction lies along the line of constant apparent
turnover scale.  Fitting to only the ``wiggles'' in Fourier space
gives weaker constraints on the distance scale which (unsurprisingly)
lie along the line of constant apparent BAO scale.

The correlation function and band-filtered correlation function yield
very similar results (with the constraints from the standard
correlation function being slightly stronger).  Their degeneracy
direction in the $(\omhh, D_V)$ parameter space lies between the two
degeneracy directions previously mentioned, implying that both the BAO
scale and correlation function shape are influencing the result.  The
slightly weaker constraint on the distance scale provided by the
band-filtered correlation function compared to the standard
correlation function is likely due to the suppression of information
on small and large scales by the compensated filter, which is designed
to reduce potential systematic errors in modelling the shape of the
clustering pattern.

For our cosmological parameter fits in the remainder of the paper we
used the standard correlation function as our default choice of
statistic.  The correlation function provides the tightest measurement
of the distance scale from our current dataset and encodes the most
significant detection of the BAO signal.

\section{Distilled parameters}
\label{secdist}

For each of the clustering statistics determined above, the
measurement of $D_V$ is significantly correlated with the matter
density $\omhh$ which controls both the shape of the clustering
pattern and the length-scale of the standard ruler (see Figure
\ref{figprobdv}).  It is therefore useful to re-cast these BAO
measurements in a manner less correlated with $\omhh$ and more
representative of the observable combination of parameters constrained
by the BAOs.  These ``distilled parameters'' are introduced and
measured in this Section.

\subsection{CMB information}
\label{seccmb}

The length-scale of the BAO standard ruler and shape of the linear
clustering pattern is calibrated by Cosmic Microwave Background data.
The cosmological information contained in the CMB may be conveniently
encapsulated by the Wilkinson Microwave Anisotropy Probe (WMAP)
``distance priors'' (Komatsu et al.\ 2009).  We use the 7-year WMAP
results quoted in Komatsu et al.\ (2011).

Firstly, the CMB accurately measures the characteristic angular scale
of the acoustic peaks $\theta_A \equiv r_s(z_*)/(1 + z_*) D_A(z_*)$,
where $r_s(z_*)$ is the size of the sound horizon at last scattering
and $D_A(z_*)$ is the physical angular-diameter distance to the
decoupling surface.  This quantity is conventionally expressed as a
characteristic acoustic index:
\begin{equation} 
\ell_A \equiv \pi/\theta_A = \pi (1+z_*)  \frac{D_A(z_*)}{r_s(z_*)} =
302.09 \pm 0.76 .
\end{equation}
The complete CMB likelihood is well-reproduced by combining this
measurement of $\ell_A$ with the ``shift parameter'' defined by
\begin{equation}
{\mathcal R} \equiv \frac{\sqrt{\om H_0^2}}{c} (1+z_*) D_A(z_*) =
1.725 \pm 0.018
\end{equation}
and the redshift of recombination (using the fitting function given as
equations 66-68 in Komatsu et al.\ 2009)
\begin{equation}
z_* = 1091.3 \pm 0.91 .
\end{equation}
The inverse covariance matrix for $(\ell_A, {\mathcal R}, z_*)$ is
given as Table 10 in Komatsu et al.\ (2011) and is included in our
cosmological parameter fit.

\subsection{Measuring $A(z)$}
\label{secaz}

\begin{figure}
\begin{center}
\resizebox{\columnwidth}{!}{\rotatebox{270}{\includegraphics{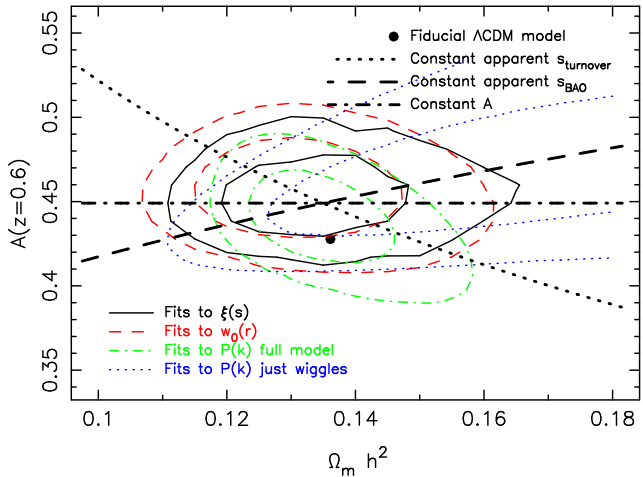}}}
\end{center}
\caption{A comparison of the results of fitting different WiggleZ
  clustering measurements in the same style as Figure \ref{figprobdv},
  except that we now fit for the parameter $A(z=0.6)$ (defined by
  Equation \ref{eqa}) rather than $D_V(z=0.6)$.}
\label{figproba}
\end{figure}

As noted in Eisenstein et al.\ (2005) and discussed in Section
\ref{secxidv} above, the parameter combination
\begin{equation}
A(z) \equiv \frac{D_V(z) \sqrt{\Omega_{\rm m} H_0^2}}{c z} ,
\label{eqa}
\end{equation}
which we refer to as the ``acoustic parameter'', is particularly
well-constrained by distance fits which utilize a combination of
acoustic oscillation and clustering shape information, since in this
situation the degeneracy direction of constant $A(z)$ lies
approximately perpendicular to the minor axis of the measured $(D_V,
\omhh)$ probability contours.  Conveniently, $A(z)$ is also
independent of $H_0$ (given that $D_V \propto 1/H_0$).  Figure
\ref{figproba} displays the measurements resulting from fitting the
parameter set $(A, \omhh, b^2)$ to the four WiggleZ clustering
statistics and marginalizing over $b^2$.  The results of the parameter
fits are displayed in Table \ref{tabres}; the correlation function
yields $A(z=0.6) = 0.452 \pm 0.018$ (i.e.\ with a measurement
precision of $4.0\%$).  For this clustering statistic in particular,
the correlation between measurements of $A(z)$ and $\omhh$ is very
low.  Given that the CMB provides a very accurate determination of
$\omhh$ (via the distance priors) we do not use the WiggleZ
determination of $\omhh$ in our cosmological parameter fits, but just
use the marginalized measurement of $A(z)$.

The measurement of $A(z)$ involves the assumption of a model for the
shape of the power spectrum, which we parameterize by $\omhh$.
Essentially the full power spectrum shape, rather than just the BAOs,
is being used as a standard ruler, although the two features combine
in such a way that $A$ and $\omhh$ are uncorrelated.  However, given
that a model for the full power spectrum is being employed, we refer
to these results as large-scale structure (``LSS'') rather than BAO
constraints, where appropriate.

\subsection{Measuring $d_z$}

In the case of a measurement of the BAOs in which the shape of the
clustering pattern is marginalized over, the $(D_V, \omhh)$
probability contours would lie along a line of constant apparent BAO
scale.  Hence the extracted distances are measured in units of the
standard-ruler scale, which may be conveniently quoted using the
distilled parameter $d_z \equiv r_s(z_d)/D_V(z)$ where $r_s(z_d)$ is
the co-moving sound horizon size at the baryon drag epoch.  In
contrast to the acoustic parameter, $d_z$ provides a purely geometric
distance measurement that does not depend on knowledge of the power
spectrum shape.  The information required to compare the observations
to theoretical predictions also varies between these first two
distilled parameters: the prediction of $d_z$ requires prior
information about $h$ (or $\omhh$), whereas the prediction of $A(z)$
does not.  We also fitted the parameter set $(d_{0.6}, \omhh, b^2)$ to
the four WiggleZ clustering statistics.  The results of the parameter
fits are displayed in Table \ref{tabres}; the correlation function
yields $d_{0.6} = 0.0692 \pm 0.0033$ (i.e.\ with a measurement
precision of $4.8\%$).  We note that, given the WiggleZ fits are in
part driven by the shape of the power spectrum as well as the BAOs,
there is a weak residual correlation between $d_{0.6}$ and $\omhh$.

When calculating the theoretical prediction for this parameter we
obtained the value of $r_s(z_d)$ for each cosmological model tested
using Equation 6 of Eisenstein \& Hu (1998), which is a fitting
formula for $r_s(z_d)$ in terms of the values of $\omhh$ and $\obhh$.
In our analysis we fixed $\obhh = 0.0226$ which is consistent with the
measured CMB value (Komatsu et al.\ 2009); we find that marginalizing
over the uncertainty in this value does not change the results of our
cosmological analysis.

We note that $r_s(z_d)$ is determined from the matter and baryon
densities in units of Mpc (not $h^{-1}$ Mpc), and thus a fiducial
value of $h$ must also be used when determining $d_z$ from data (we
chose $h = 0.71$).  However, the quoted observational result $d_{0.6}
= 0.0692 \pm 0.0033$ is actually independent of $h$.  Adoption of a
different value of $h$ would result in a shifted standard ruler scale
[in units of $h^{-1}$ Mpc] and hence shifted best-fitting values of
$\alpha$ and $D_V$ in such a way that $d_z$ is unchanged.  However,
although the observed value of $d_z$ is independent of the fiducial
value of $h$, the model fitted to the data still depends on $h$ as
remarked above.

\subsection{Measuring $R_z$}

The measurement of $d_z$ may be equivalently expressed as a ratio of
the low-redshift distance $D_V(z)$ to the distance to the
last-scattering surface, exploiting the accurate measurement of
$\ell_A$ provided by the CMB.  We note that the value of $R_z$ depends
on the behaviour of dark energy between redshift $z$ and
recombination, whereas a constraint derived from $d_z$ only depends on
the properties of dark energy at redshifts lower than $z$.  Taking the
product of $d_z$ and $\ell_A/\pi$ approximately cancels out the
dependence on the sound horizon scale:
\begin{eqnarray}
1/R_z \equiv \ell_A \, d_z/\pi &=& (1+z_*) \frac{D_A(z_*)}{r_s(z_*)}
\frac{r_s(z_d)}{D_V(z)} \label{eq:lArsDv} \\ 
&\approx& (1+z_*) \frac{D_A(z_*)}{D_V(z)} \times 1.044.  \nonumber
\end{eqnarray}
The value $1.044$ is the ratio between the sound horizon at last
scattering and at the baryon drag epoch.  Although this is a
model-dependent quantity, the change in redshift between recombination
and the end of the drag epoch is driven by the relative number density
of photons and baryons, which is a feature that does not change much
across the range of viable cosmological models.  Combining our
measurement of $d_{0.6} = 0.0692 \pm 0.0033$ from the WiggleZ
correlation function fit with $\ell_A = 302.09 \pm 0.76$ (Komatsu et
al.\ 2011) we obtain $1/R_{0.6} = 6.65 \pm 0.32$.

\subsection{Measuring distance ratios}

Finally, we can avoid the need to combine the BAO fits with CMB
measurements by considering distance ratios between the different
redshifts at which BAO detections have been performed.  Measurements
of $D_V(z)$ alone are dependent on the fiducial cosmological model and
assumed standard-ruler scale: an efficient way to measure
$D_V(z_2)/D_V(z_1)$ is by calculating $d_{z_1}/d_{z_2}$, which is
independent of the value of $r_s(z_d)$.  Percival et al.\ (2010)
reported BAO fits to the SDSS LRG sample in two correlated redshift
bins $d_{0.2} = 0.1905 \pm 0.0061$ and $d_{0.35} = 0.1097 \pm 0.0036$
(with correlation coefficient $0.337$).  Ratioing $d_{0.2}$ with the
independent measurement of $d_{0.6} = 0.0692 \pm 0.0033$ from the
WiggleZ correlation function and combining the errors in quadrature we
find that $D_V(0.6)/D_V(0.2) = 2.753 \pm 0.158$.  Percival et
al.\ (2010) report $D_V(0.35)/D_V(0.2) = 1.737 \pm 0.065$ (where in
this latter case the error is slightly tighter than obtained by adding
errors in quadrature because of the correlation between $D_V(0.2)$ and
$D_V(0.35)$).  These two distance ratio measurements are also
correlated by the common presence of $D_V(0.2)$ in the denominators;
the correlation coefficient is $0.313$.

\subsection{Comparison of distilled parameters}
\label{secdistcomp}

\begin{figure}
\begin{center}
\resizebox{\columnwidth}{!}{\includegraphics{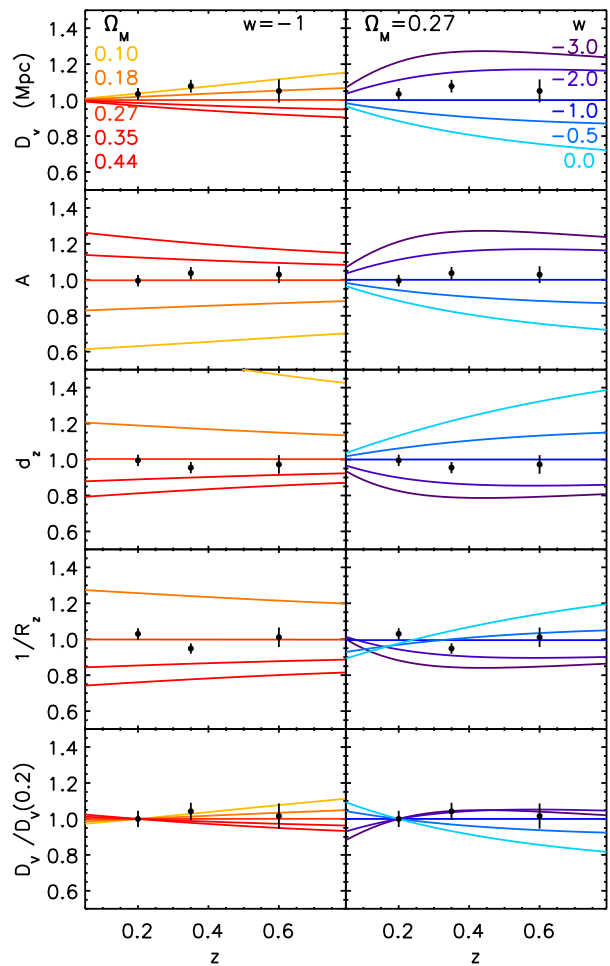}}
\end{center}
\caption{Measurements of the distilled BAO parameters extracted from
  WiggleZ and LRG datasets as a function of redshift.  From upper to
  lower we plot $D_V$ (in Mpc), the acoustic parameter $A(z)$, $d_z
  \equiv r_s(z_d)/D_V(z)$, $1/R_z \equiv \ell_A d_z/\pi$ and
  $D_V(z)/D_V(z=0.2)$.  The left-hand column shows a set of different
  cosmological models varying $\om$ with $[\ok ,w] = [0,-1]$ in each
  case.  The right-hand column displays a range of models varying $w$
  with $[\om, \ok] = [0.27,0]$ in each case.}
\label{figdv}
\end{figure}

Figure \ref{figdv} compares the measurements of the distilled
parameters introduced in this Section, using both WiggleZ and LRG
data, to a series of cosmological models varying either $\om$ or the
dark energy equation-of-state, $w$, relative to a fiducial model with
$\om = 0.27$ and $w = -1$.  These plots have all been normalized to
the fiducial model and plotted on the same scale so that the level of
information contained in the different distilled parameters can be
compared.  For presentational purposes in Figure \ref{figdv} we
converted the Percival et al.\ (2010) measurements of $d_z$ into
constraints on the acoustic parameter $A(z)$, using their fiducial
values of $r_s(z_d) = 154.7$ Mpc and $\omhh = 0.1296$ and employing
the same fractional error in the distilled parameter.

\begin{figure*}
\center
\includegraphics[width=84mm]{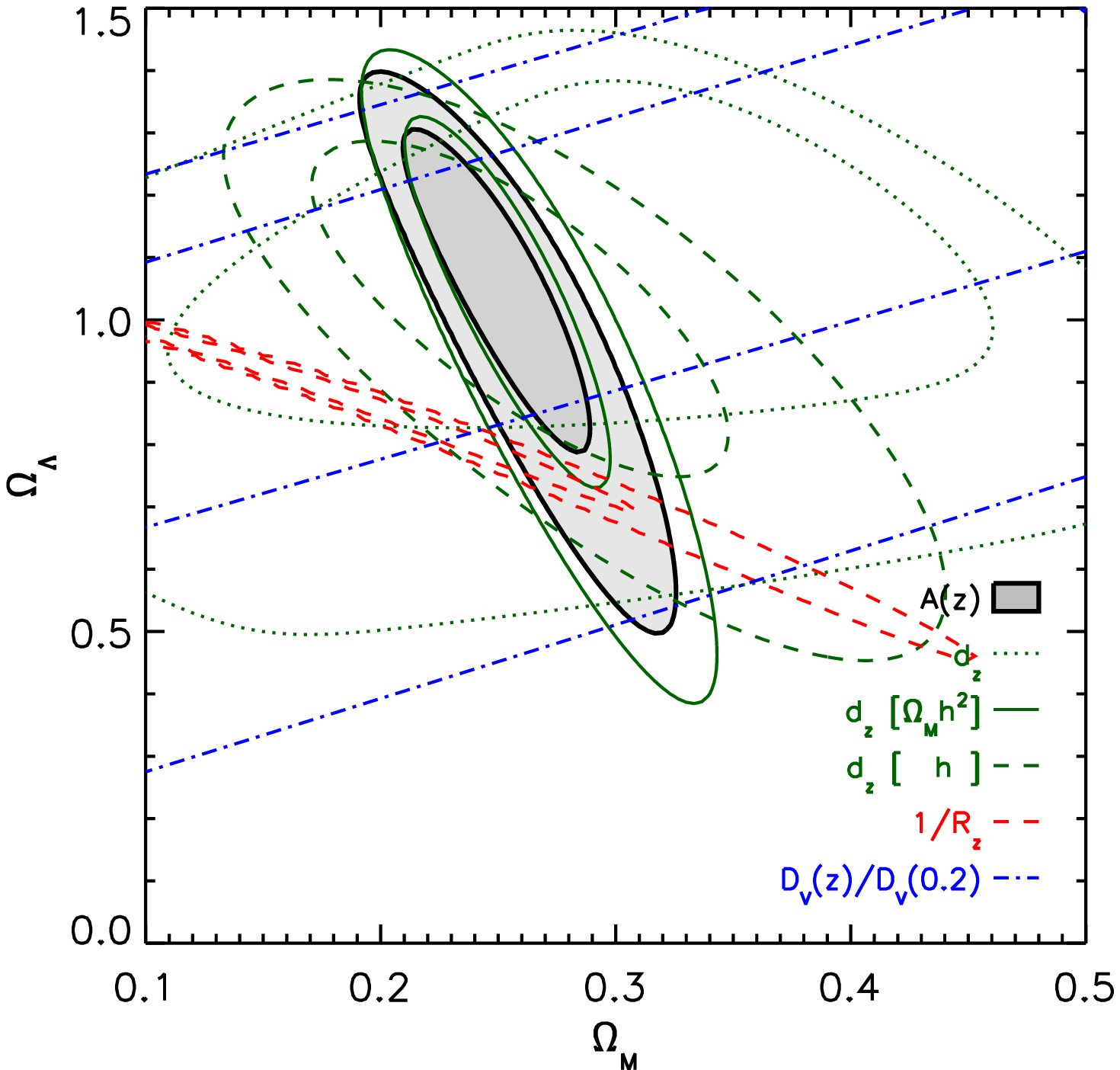}
\includegraphics[width=84mm]{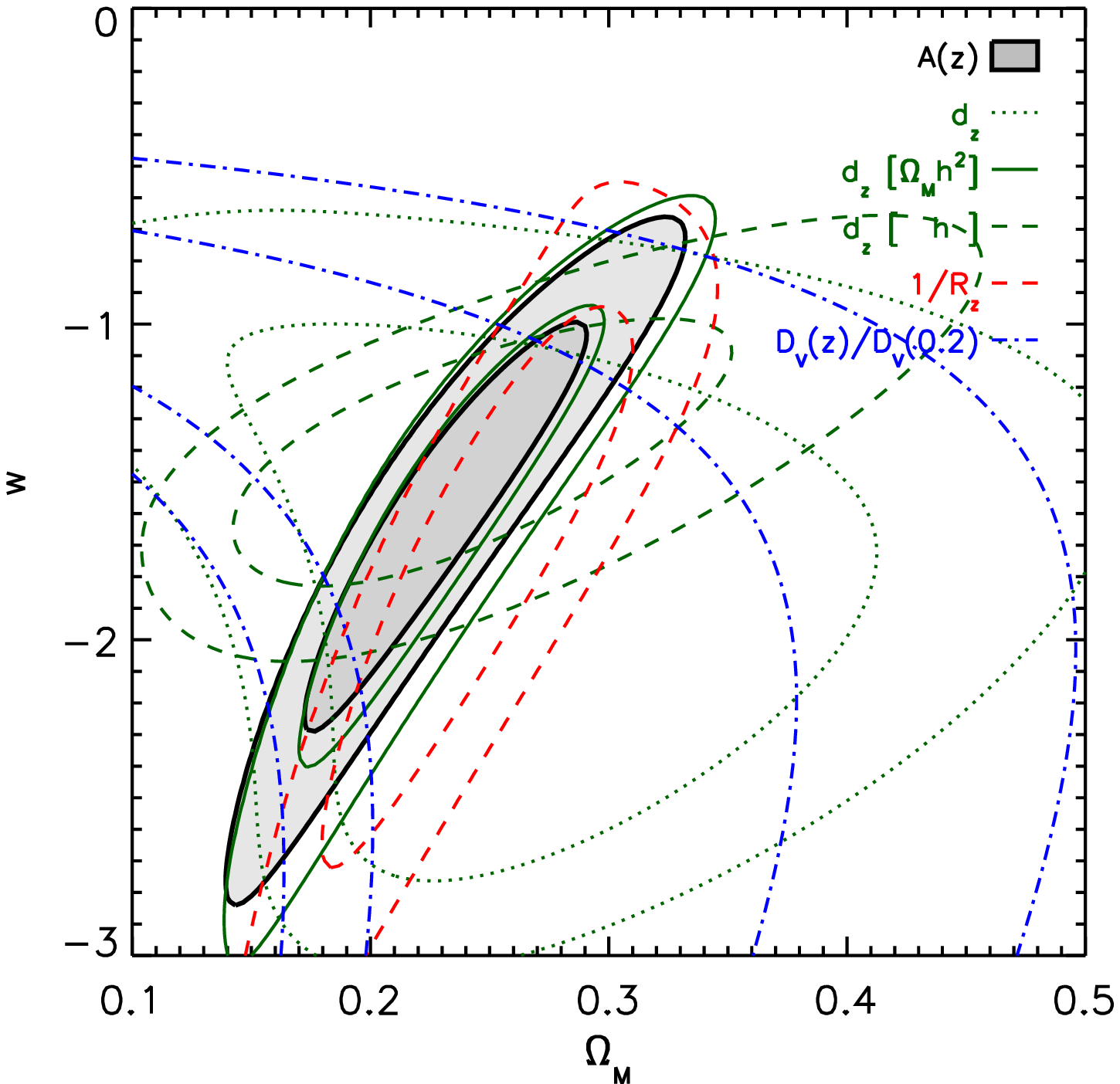}
\caption{Likelihood contours (1-$\sigma$ and 2-$\sigma$) derived from
  model fits to the different distilled parameters which may be used
  to encapsulate the BAO results from LRG and WiggleZ data.  In each
  case the contours represent the combination of the three redshift
  bins for which we have BAO data, $z=[0.2, 0.35, 0.6]$.  The thick
  solid lines with grey shading show the $A(z)$ parameter constraints,
  which are the most appropriate representation of WiggleZ data.  The
  three sets of green lines show constraints from $d_z$ assuming three
  different priors: green solid lines include a prior $\omhh = 0.1326
  \pm 0.0063$ (Komatsu et al.\ 2009, as used by Percival et
  al.\ 2010); green dotted lines marginalize over a flat prior of $0.5
  < h < 1.0$; and green dashed lines marginalize over a Gaussian prior
  of $h = 0.72 \pm 0.03$.  Red dashed lines show the $1/R_z$
  constraints, whilst fits to the ratios of BAO distances
  $D_V(z)/D_V(0.2)$ are shown by blue dash-dot lines.  The left-hand
  panel shows results for curved cosmological-constant universes
  parameterized by $(\om, \ol)$ and the right-hand panel displays
  results for flat dark-energy universes parameterized by $(\om ,w)$.
  Comparisons between these contours reveal the differing levels of
  information encoded in each distilled parameter.  By combining each
  type of distilled parameter with the CMB data in a correct manner,
  self-consistent results should be achieved.}
\label{figbaocontours}
\end{figure*}

We note that $A(z)$ and the ratio of $D_V(z)$ values are the only
distilled parameters that are independent of CMB data: the measurement
of $1/R_z$ uses $\ell_A$ from the CMB, and whilst the observed value
of $d_z$ is independent of the CMB, we need to marginalize over
$\omhh$ or $h$ in order to compare it with theoretical models.  Figure
\ref{figbaocontours} illustrates the differing information
encapsulated by each of these distilled parameters by plotting
likelihood contours of $(\om, \ol)$ for $w = -1$ and $(\om, w)$ for
$\ok = 0$ fitted to these data.  The significant differences in the
resulting likelihood contours exemplify the fact that in some cases we
have not used all the information in the galaxy data
(e.g. $D_V(z)/D_V(0.2)$ uses only the ratio of BAO scales, neglecting
information from the absolute scale of the standard ruler), while in
other cases we have already included significant information from the
CMB (e.g. $1/R_z$).  This illustrates that we must be careful to
include additional CMB data in a self-consistent manner, not
double-counting the information.  This plot also serves to illustrate
the mild tension between the BAO and CMB results, which can be seen
from the fact that the $1/R_z$ constraints are offset from the centre
of the $A(z)$ constraints.  The WiggleZ and LRG distilled BAO
parameters used in these fits are summarized for convenience in Table
\ref{tabbaores}.

\begin{table*}
\begin{center}
\caption{Measurements of the distilled BAO parameters at redshifts
  $z=0.2$, $0.35$ and $0.6$ from LRG and WiggleZ data, which are used
  in our cosmological parameter constraints.  The LRG measurements at
  $z=0.2$ and $z=0.35$ are correlated with coefficient $0.337$
  (Percival et al.\ 2010).  The two distance ratios $d_{0.2}/d_{0.35}$
  and $d_{0.2}/d_{0.6}$ are correlated with coefficient $0.313$.  The
  different measurements of $1/R_z$ are correlated by the common
  presence of the $\ell_A$ variable and by the covariance between
  $d_{0.2}$ and $d_{0.35}$.  Our default cosmological fits use the
  WiggleZ measurement of $A(z=0.6)$ combined with the LRG measurements
  of $d_{0.2}$ and $d_{0.35}$, indicated by bold font.}
\label{tabbaores}
\begin{tabular}{ccccc}
\hline
  & $A(z)$ & $d_z$ & $D_V(z)/D_V(0.2)$ & $1/R_z$ \\
  & measured & measured & $=d_{0.2}/d_z$ & $=\ell_A d_z/\pi$ \\
\hline
$z=0.2$ (LRG)
& $0.488 \pm 0.016$
& ${\mathbf 0.1905 \pm 0.0061}$
& ---
& $18.32 \pm 0.59$
\\
$z = 0.35$ (LRG)
& $0.484 \pm 0.016$
& ${\mathbf 0.1097 \pm 0.0036}$ 
& $1.737 \pm 0.065$
& $10.55 \pm 0.35$
\\
$z=0.6$ (WiggleZ)
& ${\mathbf 0.452 \pm 0.018}$
& $0.0692 \pm 0.0033$
& $2.753 \pm 0.158$
& $6.65 \pm 0.32$
\\
\hline
\end{tabular}
\end{center}
\end{table*}

\section{Cosmological parameter measurements}
\label{seccosmofit}

In this Section we present cosmological parameter fits to the
distilled BAO parameters measured above.  We consider two versions of
the standard cosmological model.  The first version is the standard
$\Lambda$CDM model in which dark energy is a cosmological constant
with equation-of-state $w = -1$ and spatial curvature is a free
parameter; we fit for $\om$ and the cosmological constant density
$\ol$.  The second version is the flat $w$CDM model in which spatial
curvature is fixed at $\ok = 0$ but the equation of state of dark
energy is a free parameter; we fit for $\om$ and $w$.

Unless otherwise stated, we fitted these cosmological models to the
WiggleZ measurement of $A(z=0.6)$ combined with the measurements of
$d_z$ from LRG samples at $z = 0.2$ and $z = 0.35$ (Percival et
al.\ 2010).  The distilled parameter $A(z)$ (measured from the galaxy
correlation function) is the most appropriate choice for quantifying
the WiggleZ BAO measurement because it is uncorrelated with $\omhh$,
as demonstrated by Figure \ref{figproba}.  The parameter $d_z$
provides the best representation of the Percival et al.\ (2010) BAO
data, because the shape of the clustering pattern was marginalized
over in that analysis.

\begin{figure*}
\center
\includegraphics[width=82mm]{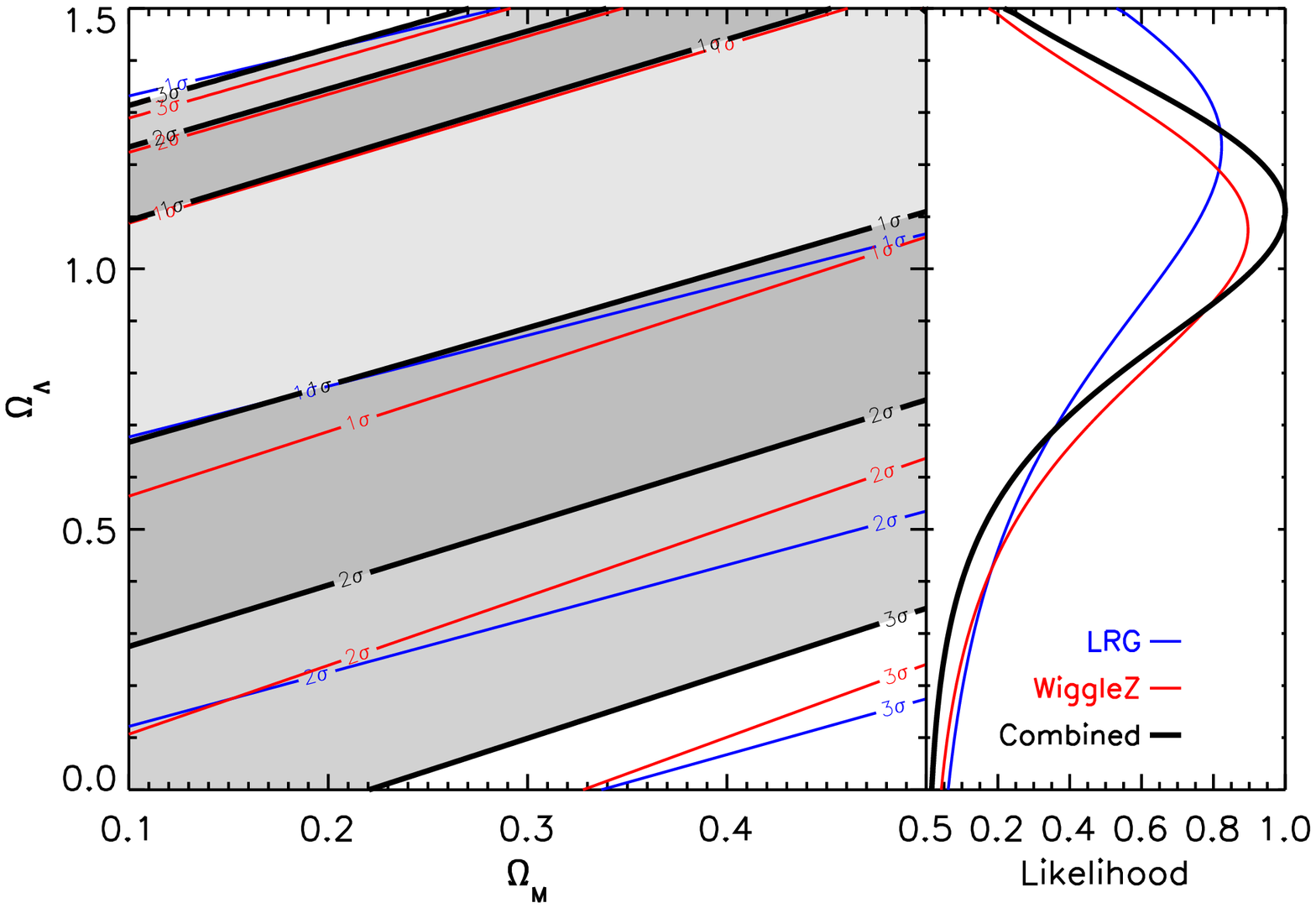}
\includegraphics[width=82mm]{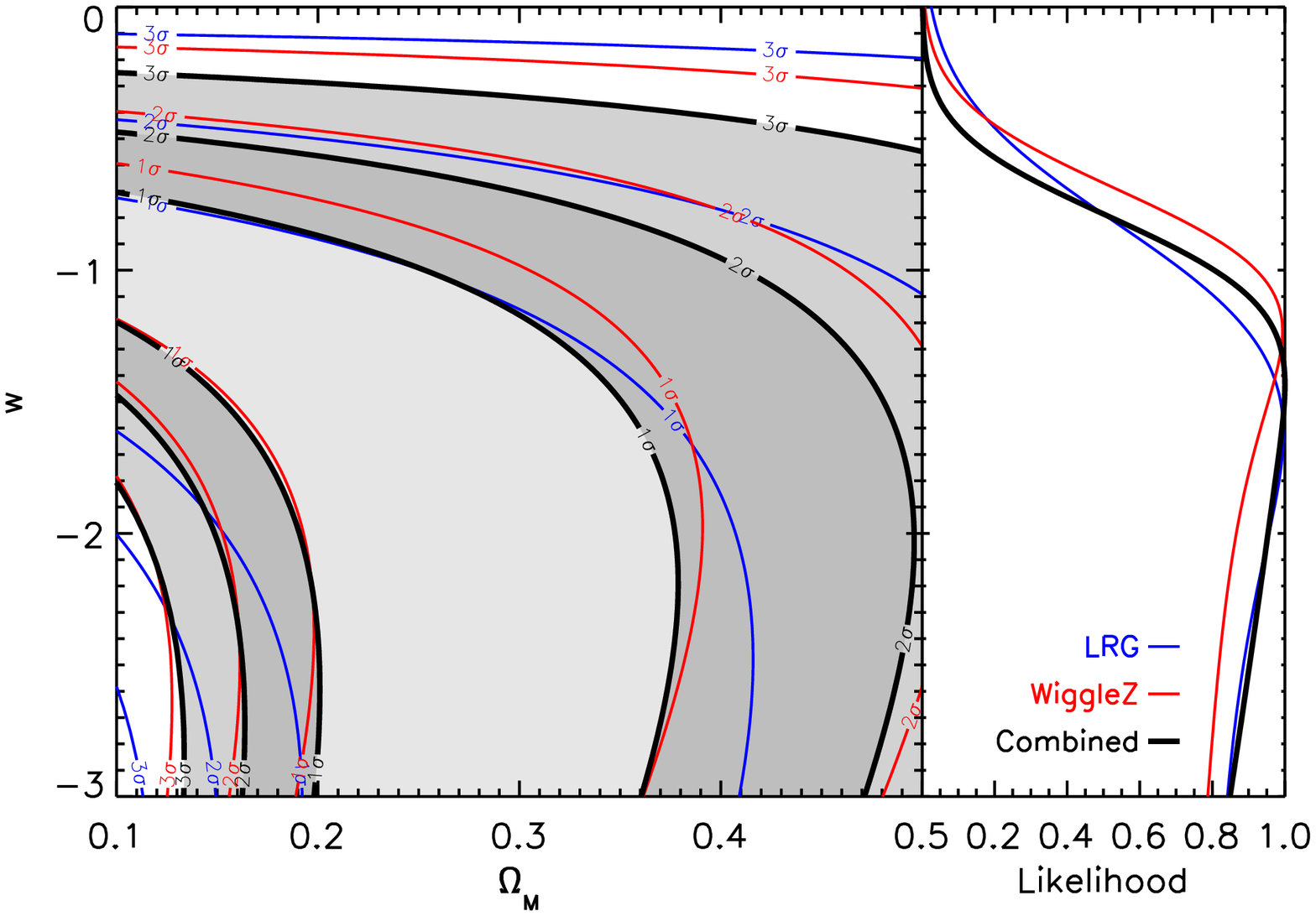}
\caption{Likelihood contours for cosmological parameter fits to BAO
  measurements using $D_V(z)$ distance ratios.  We fit two different
  models: curved cosmological-constant universes parameterized by
  $(\om, \ol)$ and flat dark-energy universes parameterized by $(\om,
  w)$.  Blue contours show the constraints using the measurement of
  $D_V(0.35)/D_V(0.2)$ obtained by Percival et al.\ (2010).  Red
  contours display the new constraints in $D_V(0.6)/D_V(0.2)$ derived
  using WiggleZ data.  The combination of these measurements is
  plotted as the grey shaded contours.  One dimensional marginalized
  likelihoods for $\ol$ and $w$ are displayed on the right-hand side
  of each contour plot.  In the flat $w$CDM model, the BAO distance
  ratios alone require accelerating expansion ($w < -1/3$) with a
  likelihood of $99.8\%$.  Compared to the LRG data, WiggleZ does not
  favour as high values of $\ol$ or as negative values of $w$.}
\label{figDvDv}
\end{figure*}

\begin{figure*}
\center
\includegraphics[width=82mm]{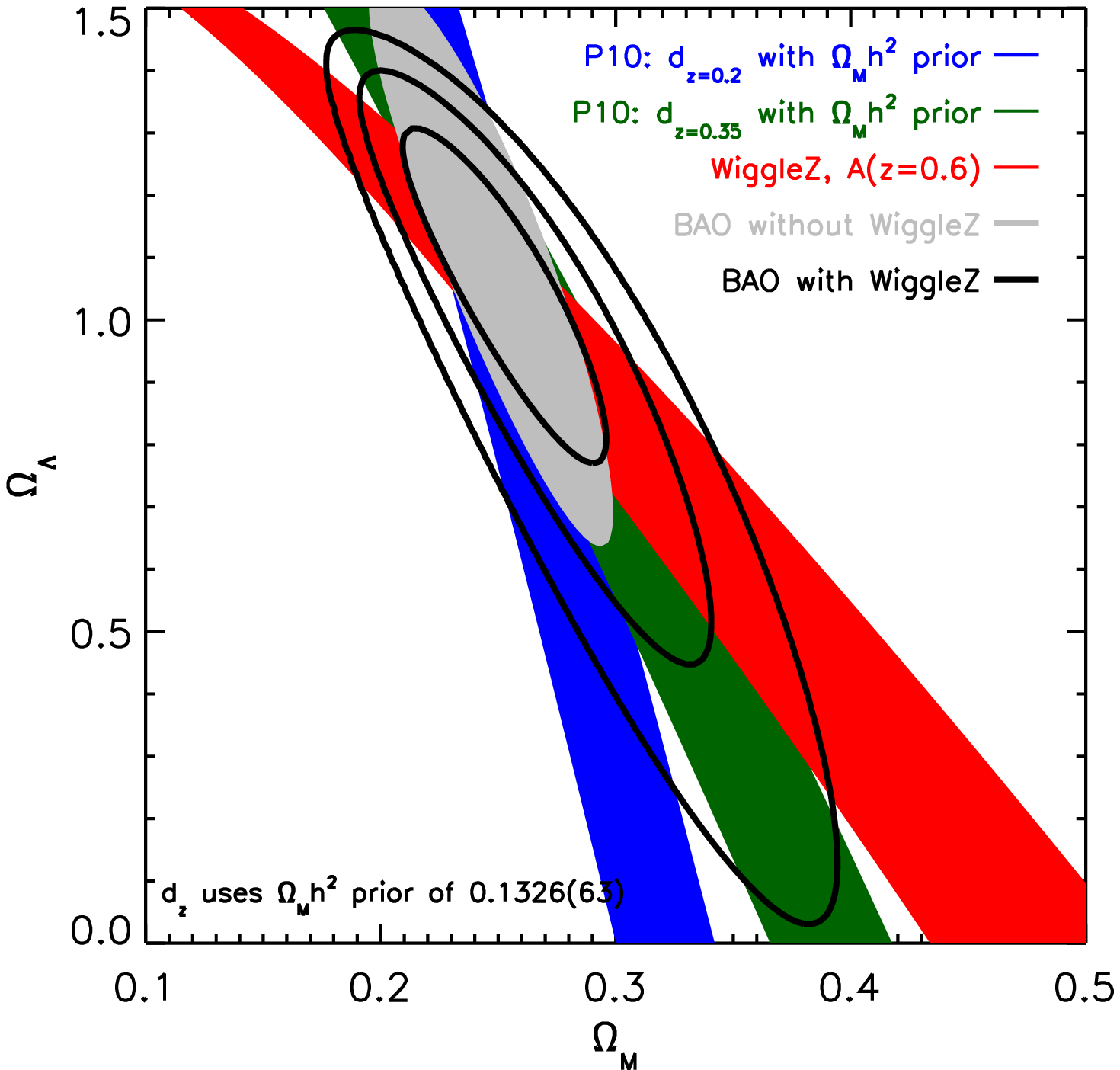}
\includegraphics[width=82mm]{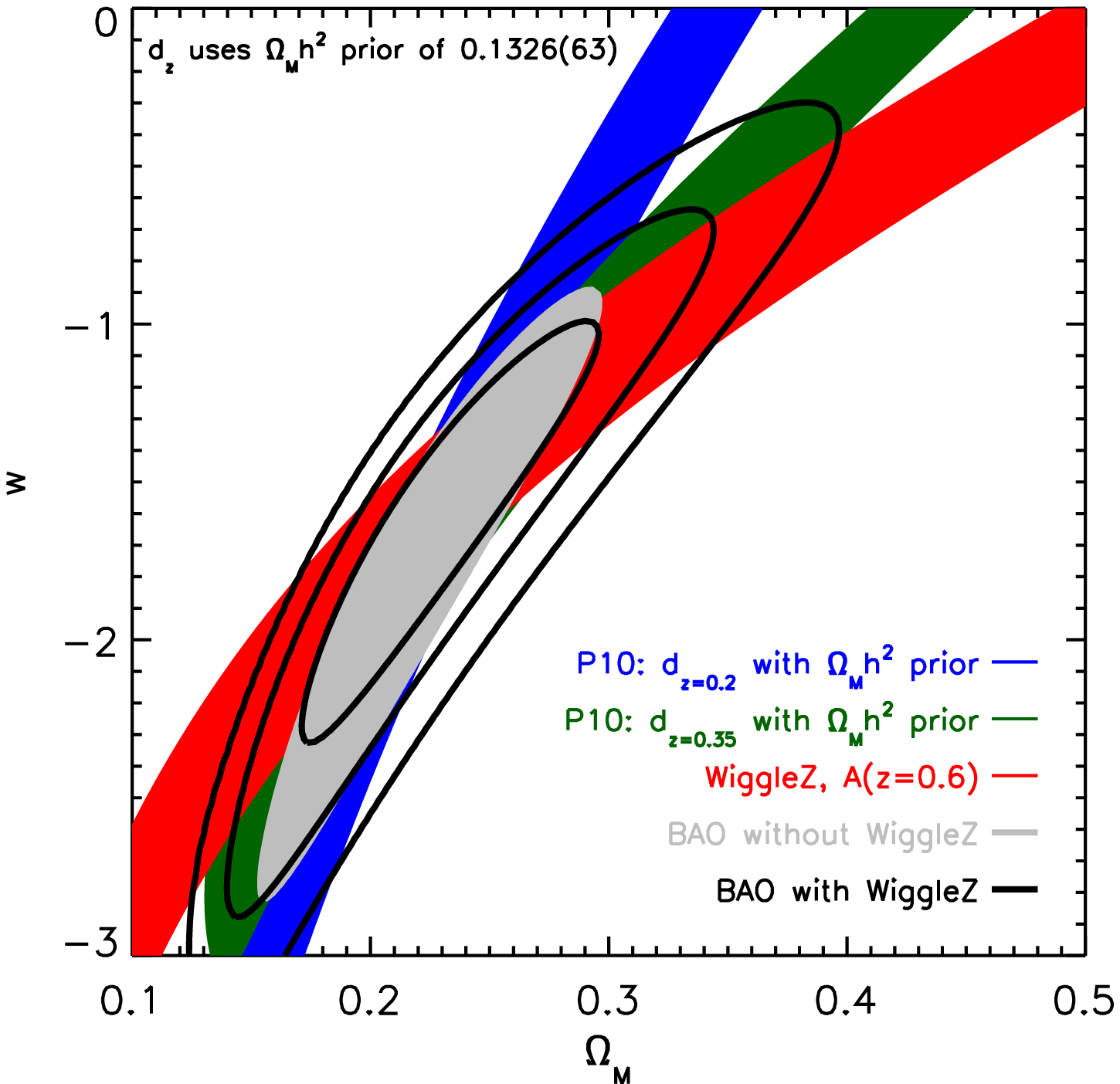}
\caption{A comparison of the WiggleZ results and previous BAO
  measurements.  The blue and green solid contours show the redshift
  $z=0.2$ and $z=0.35$ bins from Percival et al.\ (2010), using the
  $d_z$ parameter and including a CMB-motivated prior $\omhh = 0.1326
  \pm 0.0063$.  The red contour displays the fit to the WiggleZ
  measurement of $A(z)$ at $z=0.6$, which is independent of any prior
  in $\omhh$.  We note that the degeneracy directions in these plots
  rotate with redshift, demonstrating the utility of combining
  measurements at different redshifts.  The grey shaded region shows
  the combination of $d_{0.2}$ and $d_{0.35}$ (including the
  correlation).  The black solid lines are the 1, 2 and 3-$\sigma$
  contours of the final result combining all these BAO measurements.
  The addition of WiggleZ data reduces the 1-$\sigma$ uncertainty by
  about 50\% in $\ol$ and about 30\% in $w$, particularly disfavouring
  high values of $\ol$ and more negative values of $w$.  We note that
  in our final cosmological measurements we do not use a prior in
  $\omhh$ but instead combine our results with the WMAP distance
  priors.}
\label{fig:chi2bao}
\end{figure*}

\subsection{BAO alone}

As a first step in the cosmological analysis we considered the
constraints on cosmological parameters obtained using BAO data alone.
Two of the distilled parameters allow the derivation of cosmological
constraints based on only BAO measurements: the ratio of $D_V(z)$
measurements and the acoustic parameter, $A(z)$.

Ratios of $D_V(z)$ measurements are of particularly interest because
they provide constraints on the cosmic expansion history using
geometric information alone, independent of the shape of the
clustering pattern and the absolute scale of the standard ruler.
Plots of the resulting cosmological constraints in our two test models
($\Lambda$CDM and $w$CDM) are shown in Figure \ref{figDvDv}.  Using
only LRG BAO data, the single available distance ratio
$D_V(0.35)/D_V(0.2)$ provided relatively weak constraints on
cosmological parameters and, in particular, could not confirm the
acceleration of the expansion of the universe.  The addition of the
second distance ratio based on WiggleZ data, $D_V(0.6)/D_V(0.2)$,
significantly improves these constraints.  For the first time, purely
geometric distance ratios from BAO measurements demonstrate that the
cosmic expansion is accelerating: assuming the flat $w$CDM model, a
dark energy fluid with $w < -1/3$ is required with a likelihood of
$99.8\%$ (assuming the flat prior $-3 < w < 0$).

The improvement in the $\chi^2$ statistic comparing the best-fitting
$\Lambda$CDM model $(\om, \ol) = (0.25,1.1)$ to the Einstein de-Sitter
model $(\om, \ol) = (1.0,0.0)$ is $\Delta \chi^2 = 18$, whilst
comparing the best-fitting $\Lambda$CDM model to the open CDM model
$(\om, \ol) = (0.27,0.0)$ we obtain $\Delta \chi^2 = 8$.  Even given
the extra parameter in the $\Lambda$CDM model, information criteria
tests consider this level of improvement in $\chi^2$ to be significant
evidence in favour of the $\Lambda$CDM model compared to a model with
no dark energy.

Following the addition of the WiggleZ measurement, the BAO data alone
require accelerating cosmic expansion with a higher level of
statistical confidence than the initial luminosity distance
measurements from supernovae that are considered to be the first
direct evidence of accelerating expansion (compare the left-hand panel
of Figure \ref{figDvDv} with Figure 6 of Riess et al.\ 1998).
Although these BAO measurements are not yet competitive with the
latest supernova constraints, it is nevertheless reassuring that a
standard-ruler measurement of the expansion of the universe, subject
to an entirely different set of potential systematic uncertainties,
produces a result in agreement with the standard-candle measurement.

The cosmological constraints from the acoustic parameter $A(z)$ are
much more constraining than the $D_V(z)$ ratios because they
implicitly incorporate a model for the clustering pattern and standard
ruler scale as a function of $\omhh$.  Figure \ref{fig:chi2bao}
displays the resulting cosmological parameter fits to the WiggleZ
measurement of $A(z=0.6)$ combined with the LRG measurements of
$d_{z=0.2,0.35}$.  For the purposes of this Figure we combine the
$d_z$ measurements with a prior $\omhh = 0.1326 \pm 0.0063$ (following
Percival et al.\ 2010); in the next sub-section we use the WMAP
distance priors instead.  The improvement delivered by the WiggleZ
data can be seen by comparing the shaded grey contour, which is the
combination of the LRG results, with the solid black contours
representing the total BAO constraint including the new WiggleZ data.
The current WiggleZ dataset delivers an improvement of about $50\%$ in
the measurement of $\ol$ and about $30\%$ in the measurement of $w$,
based on LSS data alone.  The marginalized parameter measurements are
$\om = 0.25^{+0.05}_{-0.04}$ and $\ol = 1.1^{+0.2}_{-0.4}$ (for
$\Lambda$CDM) and $\om = 0.23 \pm 0.06$ and $w = -1.6^{+0.6}_{-0.7}$
(for $w$CDM).

The BAO results continue to prefer a more negative dark energy
equation-of-state or a higher cosmological constant density than the
CMB or supernovae data.  We explore this further in the following
Section.

\subsection{BAO combined with CMB and SNe}

\begin{figure*}
\begin{center}
\resizebox{\textwidth}{!}{\includegraphics{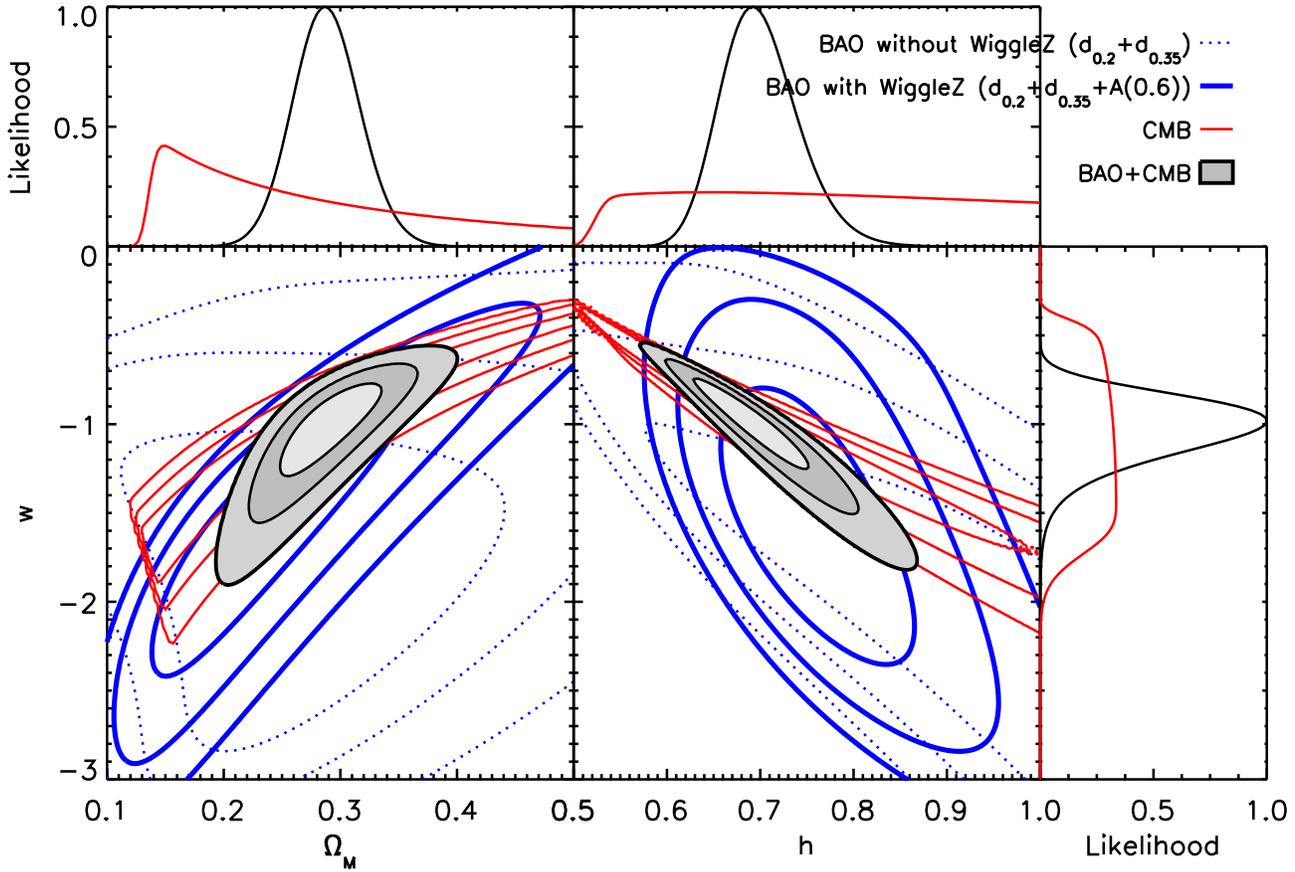}}
\end{center}
\caption{Likelihood contours for cosmological parameter fits to LRG
  and WiggleZ BAO data, compared to and combined with CMB
  measurements.  We have fitted for three parameters ($\om$, $w$ and
  $h$) assuming a flat Universe.  The left-hand panel of contours
  marginalizes over $h$ and the right-hand panel of contours
  marginalizes over $\om$.  The LSS constraints appear in blue, as
  dotted lines showing the Percival et al.\ (2010) results for $d_z$
  with no CMB prior, and as solid lines combining that data with the
  WiggleZ measurement of $A(z)$.  The red contours show the CMB
  results (using the WMAP distance priors).  The grey shaded contours
  display the combination of BAO and CMB measurements.  One
  dimensional likelihood distributions for each of the parameters are
  shown as the insets.  The BAO constraints shown here are much wider
  than those plotted in Figure \ref{fig:chi2bao} because they do not
  assume any prior in $\omhh$.  The improvement contributed by the
  BAO, as compared to the CMB constraints alone, can be seen by
  comparing the black likelihood distributions, representing the total
  constraints, with the red likelihood distributions, which are for
  the CMB alone.  (The sharp edges of the CMB likelihood distributions
  arise due to the flat priors $0.5 < h < 1.0$ and $0.1 < \om < 0.5$
  we adopted).}
\label{fig:chi2baocmb}
\end{figure*}

We now combine these large-scale structure (LSS) measurements with
other cosmological datasets.  We incorporated the CMB data using the
WMAP distance priors in $(\ell_A, {\mathcal R}, z_*)$ described in
Section \ref{seccmb}.  We fitted a model parameterized by $(\om, w,
h)$ using flat priors $0.1 < \om < 0.5$, $0.5 < h < 1.0$ and $-3 < w <
0$ and assuming a flat Universe ($\ok = 0$).  Figure
\ref{fig:chi2baocmb} compares the combined LSS cosmological parameter
measurements to the CMB constraints in both the $(\om, w)$ and $(h,
w)$ planes, marginalizing over $h$ and $\om$ respectively.  The
marginalized measurements of each parameter are $\om =
0.287^{+0.029}_{-0.028}$, $w = -0.982^{+0.154}_{-0.189}$ and $h =
0.692^{+0.044}_{-0.038}$.

We also combined these constraints with those arising from type Ia
supernovae data from the ``Union2'' compilation by Amanullah et
al.\ (2010), which includes data from Hamuy et al.\ (1996), Riess et
al.\ (1999, 2007), Astier et al.\ (2006), Jha et al.\ (2006),
Wood-Vasey et al.\ (2007), Holtzman et al.\ (2008), Hicken et
al.\ (2009) and Kessler et al.\ (2009).  Using all of these data sets
(LSS+CMB+SN) the best fitting $w$CDM model is $(\om,w) = (0.284 \pm
0.016, -1.026 \pm 0.081)$.  This model provides a a good fit to
the data with a minimum $\chi^2$ per degree of freedom of 0.95.

The best-fitting parameter values based on LSS data alone are offset
by slightly more than one standard deviation from the best-fitting
parameter values of the combined LSS+CMB+SN fit.  This mild tension
between LSS and CMB+SN results was already evident in the Percival et
al.\ (2007) measurement of BAOs in the SDSS: the LSS measurements
prefer a slightly higher rate of acceleration, or a more negative
equation of state of dark energy.  The new WiggleZ results reported
here serve to amplify that tension, although the level of discrepancy
is not statistically significant.

In order to investigate whether more complex models give a better
statistical fit to these datasets we performed fits of two additional
cosmological models.  The first is a $w$CDM model in which we allow
curvature to be a free parameter; i.e.\ we fit for $\om$, $\ok$ and
$w$.  The second is a $w(a)$CDM model in which the equation-of-state
of dark energy is allowed to vary linearly with scale factor (CPL
parameterization; Chevallier \& Polarski 2001, Linder 2003); i.e.\ we
fit for $\om$, $w_0$, and $w_a$.  If these additional parameters are
justified we should find that the minimum value of the $\chi^2$
statistic decreases significantly.  However, we in fact find that the
addition of the extra degrees of freedom improves the $\chi^2$ value
of the best-fitting model by at most $\Delta\chi^2 = 0.6$ compared to
the $\Lambda$CDM model, and significantly degrades the errors in our
fitted parameters.  In terms of information criteria this does not
represent a sufficient improvement to justify the addition of the
extra degree-of-freedom.

\section{Conclusions}

We summarize our conclusions as follows:

\begin{itemize}

\item This intermediate sample of the WiggleZ Dark Energy Survey,
  containing $N=132{,}509$ galaxies, permits a convincing detection of
  baryon acoustic oscillations at the highest redshift achieved to
  date, $z=0.6$.  We have quantified the baryon oscillations in both
  the galaxy correlation function and power spectrum statistics.  The
  statistical significance of the correlation function detection
  exceeds 3-$\sigma$.

\item We present the first measurement from a galaxy survey of the
  band-filtered BAO estimator $w_0(r)$ defined by Xu et al.\ (2010),
  which also exhibits strong evidence of BAOs, and we suggest some
  improvements in its implementation.  The distance-scale measurements
  resulting from $w_0(r)$ are slightly less precise than those
  obtained by fits to the standard correlation function, likely due to
  the suppression by the filtering function of clustering information
  on small and large scales.

\item The clustering statistics are well-fit by a non-linear power
  spectrum model including a Gaussian damping factor which models
  coherent flows on 100 $h^{-1}$ Mpc scales.  We show that our results
  are not sensitive to the details of the quasi-linear model by
  comparing five different implementations suggested in the
  literature.  We use the GiggleZ N-body simulation to demonstrate
  that scale-dependent bias effects in the WiggleZ galaxy distribution
  have an amplitude of $1\%$ on $10 \, h^{-1}$ Mpc scales, compared to
  $10\%$ for more highly-biased LRGs.

\item We measured the distance-redshift relation $D_V(z=0.6)$ with an
  accuracy of about $5\%$ using the scale dilation method.  The
  self-consistency of the measurement amongst the independent
  clustering statistics suggests that systematic measurement errors
  are not dominating these results.  Our data best constrain the
  quantity $A(z) = D_V(z) \sqrt{\Omega_{\rm m} H_0^2}/cz$, which lies
  perpendicular to the principal degeneracy direction of our contours
  and which is measured with an accuracy of $4\%$: we find $A(z=0.6) =
  0.452 \pm 0.018$.

\item The distance ratios between our measurements and previous
  analyses of the Luminous Red Galaxy distribution at redshifts
  $z=0.2$ and $z=0.35$ are consistent with a flat $\Lambda$CDM
  cosmological model which also provides a good fit to the CMB
  distance priors.  Addition of the WiggleZ data allows us to
  establish, using geometric distance ratios alone, that the
  equation-of-state of dark energy drives an accelerating expansion
  ($w < -1/3$) with $99.8\%$ likelihood, assuming a flat prior $-3 < w
  < 0$.  The current WiggleZ dataset delivers an improvement of about
  $50\%$ in the measurement of $\ol$ and $30\%$ in the measurement of
  $w$, based on BAO data alone.  The WiggleZ measurement confirms the
  mild tension that was previously reported between CMB and BAO
  measurements of the acceleration of the Universe, whereby the BAO
  data favour a slightly higher acceleration rate.

\item Cosmological parameter fits using BAO and CMB data are
  consistent with those based on current supernovae data; this
  cross-check does not yield evidence for systematic errors.
  Combining all current BAO, CMB and supernovae data we find that the
  best-fitting $w$CDM model is $(\om, w) = (0.284 \pm 0.016, -1.026
  \pm 0.081)$.  These data do not justify the addition of another
  degree-of-freedom such as non-zero spatial curvature or an evolving
  equation-of-state of dark energy.

\end{itemize}

When the complete BAO dataset from the WiggleZ survey is available, we
plan to split our data into redshift bins and explore fitting for the
BAOs separately in the tangential and radial directions, improving the
cosmological constraints presented in this paper.

The WiggleZ dataset enables further tests of the cosmological model,
complementary to those involving baryon acoustic oscillations.  Two
such results will be published at a similar time as the current paper.
Firstly, we have used redshift-space distortions in the WiggleZ sample
to measure accurately the growth rate of structure in the redshift
range $0.1 < z < 0.9$, finding that this is consistent with the
predictions of the $\Lambda$CDM model (Blake et al.\ 2011a).
Secondly, we have used the Alcock-Paczynski test to perform
non-parametric reconstructions of the cosmic expansion history (Blake
et al.\ 2011b).

\section*{Acknowledgments}

We acknowledge financial support from the Australian Research Council
through Discovery Project grants funding the positions of SB, MP, GP
and TD.  SMC acknowledges the support of the Australian Research
Council through a QEII Fellowship.  MJD thanks the Gregg Thompson Dark
Energy Travel Fund for financial support.

GALEX (the Galaxy Evolution Explorer) is a NASA Small Explorer,
launched in April 2003.  We gratefully acknowledge NASA's support for
construction, operation and science analysis for the GALEX mission,
developed in co-operation with the Centre National d'Etudes Spatiales
of France and the Korean Ministry of Science and Technology.

Finally, the WiggleZ survey would not be possible without the
dedicated work of the staff of the Australian Astronomical Observatory
in the development and support of the AAOmega spectrograph, and the
running of the AAT.

\end{document}